\documentclass[12pt,useAMS, usenatbib]{mnras}
\usepackage{amsmath}
\usepackage{amssymb}
\usepackage{graphicx}
\usepackage{subfigure}
\usepackage{bm}

\newcommand{\be}{\begin{equation}} 
\newcommand{\ee}{\end{equation}}

\newcommand{\ba}{\begin{eqnarray}} 
\newcommand{\ea}{\end{eqnarray}}

\newcommand{\mjup}{M_{\rm J}}
\newcommand{\rjup}{R_{\rm J}}

\newcommand{\msun}{M_{\odot}}
\newcommand{\rsun}{R_{\odot}}

\newcommand{\ain}{a_{\rm in}}
\newcommand{\aout}{a_{\rm out}}

\newcommand{\ein}{e_{\rm in}}
\newcommand{\eout}{e_{\rm out}}
\newcommand{\Imut}{I_{\rm mut}}

\newcommand{\fppfirst}{f_{\rm pp,1p}^{(1)}}
\newcommand{\fpsfirst}{f_{\rm ps,1p}^{(1)}}
\newcommand{\fejfirst}{f_{\rm ej,1p}^{(1)}}

\newcommand{\fppsec}{f_{\rm pp,1p}^{(2)}}
\newcommand{\fpssec}{f_{\rm ps,1p}^{(2)}}
\newcommand{\fejsec}{f_{\rm ej,1p}^{(2)}}

\newcommand{\bcoll}{b_{\rm coll}}

\newcommand{\fiducial}{{\texttt{fiducial}}}
\newcommand{\fidkthree}{{\texttt{fiducial-K-3}}}
\newcommand{\fidkfive}{{\texttt{fiducial-K-5}}}
\newcommand{\eqmass}{{\texttt{near-eq-mass}}}
\newcommand{\lognorm}{{\texttt{lognorm-mass}}}
\newcommand{\fourp}{{\texttt{4-planets}}}

\newcommand{\rebound}{\textsc{rebound}}
\newcommand{\reboundx}{\textsc{reboundx}}

\def\go{\mathrel{\raise.3ex\hbox{$>$}\mkern-14mu
             \lower0.6ex\hbox{$\sim$}}}

\def\lo{\mathrel{\raise.3ex\hbox{$<$}\mkern-14mu
             \lower0.6ex\hbox{$\sim$}}}
\begin{document} 

\title[In-Situ Scattering of Warm Jupiters]{In-Situ Scattering of Warm Jupiters and Implications for Dynamical Histories} 
\pagerange{\pageref{firstpage}--\pageref{lastpage}} \pubyear{2019}

\label{firstpage}

\author[K. R. Anderson, D. Lai, and B. Pu]{Kassandra R. Anderson$^{1}$\thanks{E-mail:
    kassandra@princeton.edu}\thanks{Current address: Department of Astrophysical Sciences, Princeton University, Princeton, NJ 08544}, Dong Lai$^{1,2}$, \& Bonan Pu$^{1}$  \\ \\ $^{1}$Cornell Center for Astrophysics and Planetary Science, Department of Astronomy, Cornell University, Ithaca, NY 14853, USA \\ $^{2}$Tsung-Dao Lee Institute, Shanghai Jiao Tong University, Shanghai 200240, China} 

\maketitle

\begin{abstract}
Many warm Jupiters (WJs) have substantial eccentricities, which are linked to their formation and migration histories.  This paper explores eccentricity excitation of WJs due to planet-planet scattering, beginning with 3-4 planets in unstable orbits, with the innermost planet placed in the range $(0.1 - 1)$ AU.  Such a setup is consistent with either in-situ formation or arrival at sub-AU orbits due to disk migration.  Most previous N-body experiments have focused on ``cold'' Jupiters at several AU, where scattering results in planet ejections, efficiently exciting the eccentricities of surviving planets.  In contrast, scattering at sub-AU distances results in a mixture of collisions and ejections, and the final eccentricities of surviving planets are unclear.  We conduct scattering experiments for a range of planet masses and initial spacings, including the effect of general relativistic apsidal precession, and systematically catalogue the scattering outcomes and properties of surviving planets.  A comparable number of one-planet and two-planet systems are produced. Two-planet systems arise exclusively through planet-planet collisions, and tend to have low eccentricities/mutual inclinations and compact configurations.  One-planet systems arise through a combination of ejections and collisions, resulting in higher eccentricities.  The observed eccentricity distribution of solitary WJs (lacking detection of a giant planet companion) is consistent with roughly $60 \%$ of the systems having undergone in-situ scattering, and the remaining experiencing a quiescent history. 

\end{abstract}


\section{Introduction}
The origin of warm Jupiters (WJs, giant planets with orbital periods in the range of $\sim 10 - 300$ days) remains an unsolved problem in exoplanetary science.  Whether WJs previously migrated from farther out, are currently in the process of migration, or formed in-situ, is uncertain.  In-situ formation of hot Jupiters (HJs) and other close-in planets was considered previously by \cite{lee2014,batygin2016, boley2016,lee2016}. Giant planet migration has been extensively studied in the context of HJ formation, and comes in two types. One is disk migration, in which planets are transported inwards due to torques from the protoplanetary disk \citep[e.g.][]{lin1996,tanaka2002,kley2012,baruteau2014}.   The second is high-eccentricity migration, in which the planet's eccentricity is excited to an extreme value by a stellar or planetary companion(s), so that tides raised on the planet at pericenter passages shrink and circularize the orbit.  High-eccentricity migration comes in several distinct flavors, depending on the details of the eccentricity excitation. Possibilities include excitation from an inclined companion due to Lidov-Kozai cycles \citep{lidov1962,kozai1962} or other secular perturbations \citep{wu2003,fabrycky2007,correia2011, naoz2012,petrovich2015lk,petrovich2015co,anderson2016,munoz2016,hamers2017,vick2019}, planet-planet scatterings, possibly combined with secular interactions \citep{rasio1996,nagasawa2008,nagasawa2011,beauge2012}, and secular chaos \citep{lithwick2011,lithwick2014,teyssandier2019}.  See also \cite{dawson2018} for a review.

WJs are observed to have a wide range of eccentricities.  The radial velocity WJs from the Exoplanet Orbit Database\footnote{exoplanets.org, accessed July 4, 2019.} (with minimum planet masses above $0.3 \mjup$ and semi-major axes in the range $0.1-1$ AU) have an average eccentricity $e \simeq 0.24$, and $30 \%$ of the planets have $e > 0.3$, with a maximum $e \simeq 0.93$.  Theories of planet formation and migration must be able to account for these observations.

Several mechanisms have been proposed in exciting WJ eccentricities.  One seemingly natural explanation is high-eccentricity migration. In this scenario, WJs are caught in the act of inward migration, eventually to become HJs in circular orbits.  In any high-eccentricity migration theory, a small pericenter distance is required, so that tidal dissipation may shrink the orbit within the lifetime of the host star. The majority of observed WJs have pericenter distances too large to allow for efficient tidal dissipation.  These WJs may be undergoing secular eccentricity oscillations driven by an external companion, so that they are observed during a low-eccentricity phase of the secular cycle \citep{dong2014,dawson2014,petrovich2016}.  This scenario requires a relatively close or massive companion, so that the secular perturbations that lead to eccentricity oscillations are not suppressed by general relativistic apsidal precession.   

As a planet is transported inward via high-eccentricity migration, it eventually becomes decoupled from the companion. Unless the perturber is particularly ``strong'' (i.e. close or massive), this decoupling has usually occurred by the time the migrating planet reaches WJ territory, causing the eccentricity to freeze at the maximum value. With a small (fixed) pericenter distance, the planet migrates rapidly into HJ territory, with little time spent in the WJ phase \citep[see, e.g., Fig.~1 of][]{anderson2016}\footnote{When dynamical (chaotic) tides are incorporated into the high-eccentricity migration model, highly eccentric ($e \gtrsim 0.9$) WJs form rapidly, and subsequently circularize to become HJs \citep{vick2019}.}.  Consequently, forming WJs that are likely to be observed requires a somewhat specific initial architecture. The perturber must be strong enough to drive eccentricity oscillations at sub-AU distances, but weak enough to ensure dynamical stability of the initial configuration.  

Indeed, HJ population syntheses that have chosen a broad range of properties for the perturber do not produce any appreciable numbers of WJs \citep{petrovich2015lk,anderson2016,hamers2017}.  Observationally, the number of WJs exceeds the number of HJs.  The study by \cite{petrovich2016} attempts to replicate the observed relative numbers of HJs and WJs by choosing as their initial condition a population of planets initially located at $1$ AU with strong perturbers in a narrow range of semi-major axis, located at $5-6$ AU.   However, these initial conditions still produce too few WJs compared to HJs.

Another difficulty of high-eccentricity migration in the context of WJ formation is the fact that unlike HJs, a large fraction of WJs have one or more close, low-mass neighbors \citep{huang2016}.  Such a configuration is difficult to produce in a violent high-eccentricity migration scenario. Furthermore, \cite{antonini2016} examined the subset of observed WJs with characterized (in terms of mass, semi-major axis and eccentricity) external giant planet companions, and found that most systems are inconsistent with a traditional high-eccentricity migration origin (where the WJ formed beyond 1 AU), due to the initial configurations being dynamically unstable.

These difficulties of high-eccentricity migration in reproducing properties of WJs indicate that in-situ formation or disk migration may be responsible for forming many, if not most WJs. However, both in-situ formation and disk migration by themselves have difficulties in producing eccentric WJs.  As a result, mechanisms for exciting eccentricity are needed. Planet-disk interactions are capable of exciting eccentricities, but are limited to modest values $\lesssim 0.2$ \citep{goldreich2003,tsang2014,duffell2015,ragusa2018}.  Recently, \cite{petrovich2019} studied a mechanism for transferring eccentricity from an outer planet to a WJ during the dispersal of a massive protoplanetary disk.

In \cite{anderson2017}, we considered the possibility that eccentric WJs arise due to secular perturbations from a distant companion, without requiring that the WJs be undergoing high-eccentricity migration.  Eccentricity may be excited by a highly inclined perturber via Lidov-Kozai cycles, or by an eccentric coplanar perturber due to an apsidal precession resonance.  Taking the sample of WJs with external planetary companions with characterized orbits, and assuming that the WJ formed in a circular orbit, we found that relatively high mutual inclinations are needed ($\sim 50^\circ-60^\circ$) to generate the observed eccentricities.  This finding is intriguing, and consistent with previous evidence for high inclinations in many of the same systems found by \cite{dawson2014}.  However, generating such inclinations is non-trivial, and requires an early scattering event in the system's history.

Planet-planet scattering itself remains another possibility in producing eccentric WJs \citep{mustill2017,frelikh2019}, or scattering in combination with dynamical tides \citep{marzari2019}.   A substantial literature of giant planet scattering work exists, ranging from scattering of two planets to ten or more planets \citep[e.g.][]{chambers1996,lin1997,ford2001,adams2003,chatterjee2008,ford2008,juric2008,nagasawa2011}, but most studies have focused on ``cold Jupiters,'' giant planets located at several AU.  Giant planet scattering close to the host star is less explored. 

Scattering outcomes depend on the ``Safronov number'', the squared ratio of the escape velocity from the planetary surface to the planet's orbital velocity.  When the Safronov number is much less than unity, close encounters between planets result in planet-planet collisions, with the collision product having a low eccentricity.  When the Safronov number is much greater than unity, planet ejections are expected, efficiently raising the eccentricities of the remaining planets.  WJs have Safronov numbers of order unity and thus lie in a regime in which a combination of collisions and ejections may occur, so that the degree of eccentricity excitation is unclear.  \cite{petrovich2014} undertook a scattering study of primarily HJs, with close initial spacings, and found inefficient eccentricity excitation due to a preponderance of collisions.  Whether this finding holds for WJs with a wider range of initial spacing has yet to be thoroughly explored.
  
This paper presents a systematic study of planet-planet scattering for systems of closely-spaced WJs starting with initially low eccentricities.  This setup is consistent with either in-situ formation or arrival at a sub-AU orbit by disk migration.   The goal of this paper is two-fold: (1) On observationally-motivated grounds, we aim to identify to what extent planet scattering may be contributing to eccentric WJs; (2) On the theoretical side, we aim to catalogue the scattering outcomes for planets in the range $0.1-1$ AU, where a rich variety of collisions and ejections are expected.  We conduct N-body scattering experiments of three or four giant planets, including the effect of general relativistic apsidal precession.  We explore a variety of choices for planet masses and initial spacings, determine the branching ratios for various outcomes, and analyze the properties of the ``remnant'' systems after scattering. Radial velocity observations have yielded samples of solitary WJs and WJs with an external giant planet companion, with eccentricity measurements for both samples.  We compare the results of the scattering experiments with the observed system properties.

This paper is organized as follows.  In Section \ref{sec:scattering}, we describe the setup of our N-body calculations, present branching ratios, demonstrate how the scattering results depend on various parameters, and discuss the properties of the surviving planetary systems. In Section \ref{sec:obs} we compare the results of Section \ref{sec:scattering} with observations.  We present our conclusions in Section \ref{sec:conclusion}.

\section{Scattering Experiments}\label{sec:scattering}
\subsection{Setup \& Canonical Parameters}
We begin with a system of three planets, with masses $m_1$, $m_2$, $m_3$, orbiting a host star with mass $M_\star = 1 \msun$ and radius $R_\star = 1 \rsun$.  For each planet, we sample the initial eccentricities uniformly in the range $[0.01,0.05]$, inclinations (relative to an arbitrary reference frame) uniformly in the range $[0^\circ,2^{\circ}]$, and argument of pericenter, longitude of ascending node, and mean anomaly uniformly in $[0,2 \pi]$. The initial semi-major axes are specified in units of mutual Hill radius $R_{\rm H,mut}$, so that 
\be
a_{i} - a_{i - 1} = K R_{\rm H,mut},
\label{eq:aspace}
\ee 
where
\be
R_{\rm H,mut} = \frac{ a_{i -1} + a_{i}}{2} \bigg( \frac{m_{i - 1} + m_{i}}{3 M_\star} \bigg)^{1/3}.
\label{eq:Rmut}
\ee
We adopt a fiducial spacing of $K = 4$.  The innermost semi-major axis $a_1$ is sampled uniformly in the range $[0.1 - 1]$AU.  The planet masses are chosen to be $0.5$, $1$ and $2$ $\mjup$, with randomly assigned ordering.  We draw a sample of over 3000 systems with these parameters and evolve using N-body integrations. This set of simulations constitutes our fiducial sample, which we will refer to as \fiducial\ (see also Table \ref{table1}).  Sections \ref{sec:K}, \ref{sec:a1} and \ref{sec:masses} explore how the results depend on $K$, $a_1$, and planet masses, and Section \ref{sec:fourp} considers scattering of four planets.

The choice of these initial conditions is broadly consistent with an initial ensemble of sub-AU giant planets that is stabilized under the presence of a gaseous disk, and later undergoes instabilities following disk dispersal. The ranges for the initial eccentricities and inclinations are chosen to be small, but non-zero, reflecting that small but finite eccentricities and inclinations are expected under some circumstances (e.g. by planet-disk interactions). Furthermore, these ranges, as well as our choice to space the orbits in fixed units of mutual Hill radius, are similar to previous ``cold Jupiter" N-body scattering experiments, allowing for comparison between this paper and previous work.

The N-body calculations are performed using \rebound\ \citep{rein2012}. We include the effects of apsidal precession due to general relativity using the \texttt{gr-potential} option in \reboundx \footnote{https://github.com/dtamayo/reboundx}.  When the separation between any two bodies becomes less than the sum of their radii, we assume the bodies merge, conserving mass and momentum, as in the built-in \rebound\ collision routine (see also Appendix \ref{sec:collision}).  Since we consider young giant planets, we set the radius of each planet to $R_p = 1.6 \rjup$.  Planets are considered ejected if the distance from the host star exceeds $1000$ AU, and are subsequently removed from the simulation.

We integrate systems of three unstable planets using the IAS15 integrator in \rebound\ \citep{rein2015ias15} for a timespan of $10^6 P_1$, with $P_1$ the initial orbital period of the innermost planet.  We refer to this initial, highly unstable phase (with close encounters eventually resulting in collisions or ejections) as ``Phase 1'' of the integration.  After this phase nearly $100 \%$ of the three-planet systems have become destabilized due to collisions or ejections. We continue to integrate the remaining two-planet systems for a timespan of $10^8 P_{\rm in}$ (or until another collision or ejection has occurred), where $P_{\rm in}$ is the orbital period of the inner planet at the end of ``Phase 1''.  We refer to this longer-term integration as ``Phase 2,'' and use the hybrid integrator Mercurius.  Mercurius utilizes a symplectic Wisdom-Holman integrator WHFAST for large separations between planets \citep{rein2015whfast, wisdom1991}, switching to IAS15 when the separation between any two bodies becomes less than a critical value.  We choose this critical value to be $5$ Hill radii.  The timestep for the Wisdom-Holman integrator is chosen to be $0.02 P_{\rm in}$.  Repeating a subset of the \fiducial\ sample with a timestep $0.01 P_{\rm in}$ yielded statistically identical results. 

\subsection{Scattering Outcomes} \label{sec:scatoutcome}

\begin{figure*}
\centering 
\includegraphics[width=\textwidth]{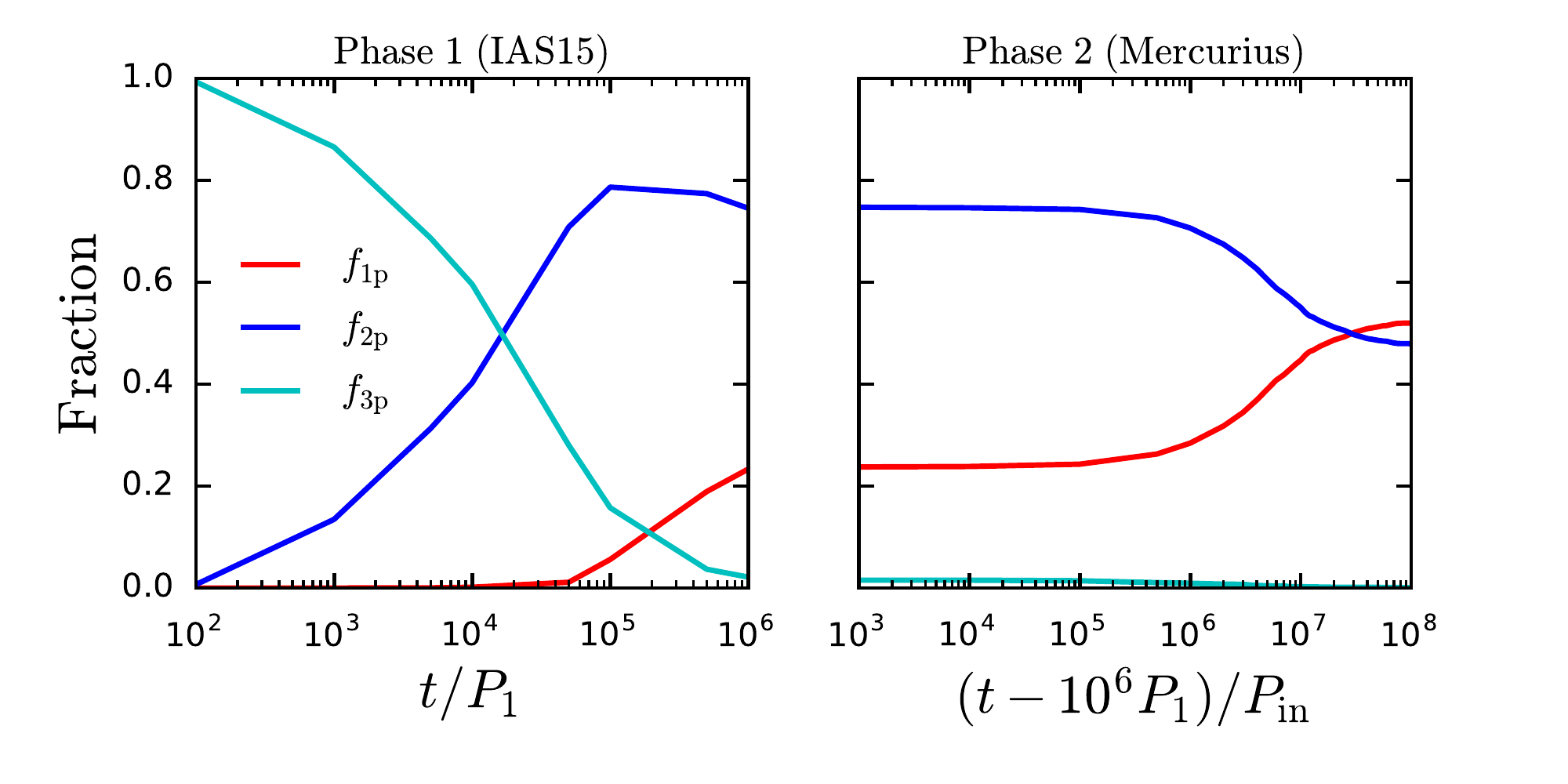}
\caption{Fraction of one, two, and three-planet systems as a function of time for the \fiducial\ set of simulations.  {\it Left}: ``Phase 1'' of the integration, in which the initial  three-planet system was evolved using the IAS15 integrator in \rebound. After $10^6$ initial orbital periods of the innermost planet have elapsed, nearly all of the three-planet systems have become destabilized, due to a combination of planet collisions and ejections. {\it Right:} ``Phase 2'' of the integration, in which we evolved the remaining two-planet systems using the hybrid Mercurius integrator, for a timespan of $10^8$ orbital periods of the inner planet of the two-planet systems.  At the end of the integration, the fractions of one and two-planet systems approach constant values.}
\label{fig:fraction}
\end{figure*}

Figure \ref{fig:fraction} shows the fractions of one, two, and three-planet systems as a function of time.  The left panel shows ``Phase 1'' of the integration using IAS15.  Since the initial systems are highly unstable, the fraction of three-planet systems quickly decays, eventually reaching a negligible value after $10^6$ initial orbital periods of the innermost planet.  During the long-term follow-up integration of the two-planet systems (``Phase 2'', right panel of Fig.~\ref{fig:fraction}), the fraction of two-planet systems declines, eventually approaching a constant value after $10^8$ orbits. The vast majority of the two-planet systems are undergoing secular interactions at this point, with constant semi-major axes and oscillating eccentricities and inclinations.  We thus conclude that the majority of the remaining 2-planet systems are stable at this time.  

We note that a number of these two-planet systems could undergo instabilities over much longer timescales.  Several empirical stability criteria for two-planets in eccentric/inclined orbits have been proposed \citep[e.g.][]{mardling2001,petrovich20152p}, but they are only reliable sufficiently far from the ``fuzzy'' stability boundary.  Using the \cite{petrovich20152p} stability criterion, we find that over $90 \%$ of our \fiducial\ two-planet systems are classified as unstable, but fall in the uncertain regime of parameter space near the boundary. Thus, the \cite{petrovich20152p} stability criterion is overly conservative in this application.  In addition, his stability criterion does not consider the potentially stabilizing effects of GR apsidal precession, as well as the possibility of planet-planet collisions.  As a result, full numerical integrations over long timescales are needed. To evaluate how likely instabilities may occur over much longer timescales, we randomly choose 30 of the closely-spaced two-planet systems (with $\aout / \ain < 3$) from the \fiducial\ sample, and integrate for an additional $10^9$ orbits of the inner planet, i.e. an order of magnitude longer.  Of these 30 systems, only 3 became destabilized.  We note that $10^9$ orbits at typical WJ distances may still  be relatively short compared to the lifetimes of observed systems.  We thus conclude that at least $\sim 10 \%$ of the two-planet systems are expected to eventually go unstable. 

\begin{figure*}
\centering 
\includegraphics[width=0.8\textwidth]{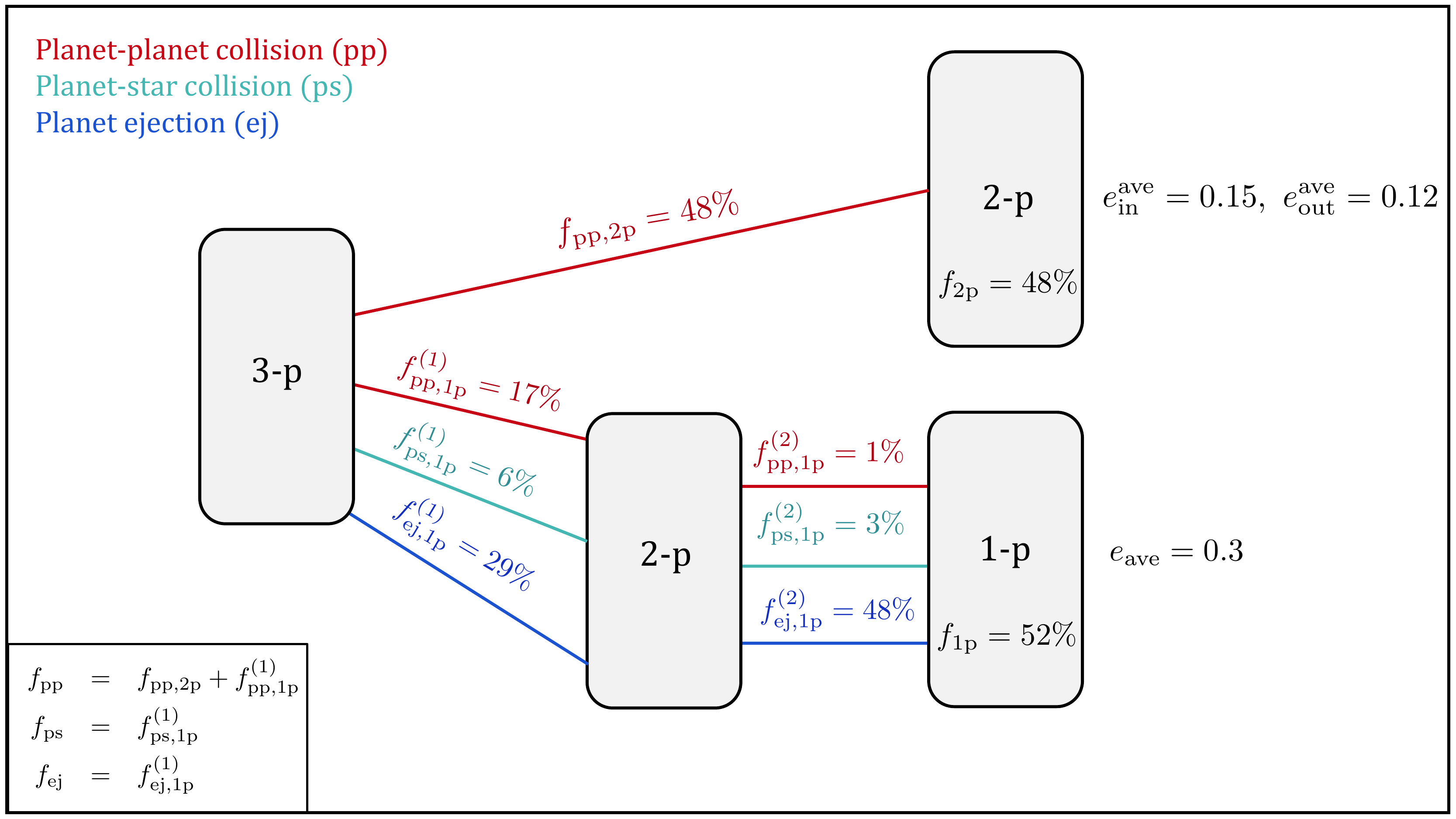}
\caption{Branching ratios for the \fiducial\ simulations, illustrating the ``decay'' of the unstable three planet-planet systems into one-planet and two-planet systems.  We also display the final average planet eccentricities.  Note the relations $f_{\rm pp,2p} + \fppfirst + \fpsfirst + \fejfirst = 100 \%$ and $f_{\rm pp,2p} + \fppsec + \fpssec + \fejsec = 100 \%$.  All stable two-planet systems arise due to planet-planet collisions, so that the possible branching ratios $f_{\rm ps, 2p} = f_{\rm ej,2p} = 0$, and are not depicted in the diagram (see also Table \ref{table_2p}).}
\label{fig:bratio}
\end{figure*}

The fractions of one and two-planet systems at the end of Phase 2 are $f_{\rm 1p} = 52 \%$ and $f_{\rm 2p} =  48 \%$ respectively.  Figure \ref{fig:bratio} shows the chain of events and branching ratios that produced these one and two-planet systems.  A close encounter can result in a planet ``loss'' due to a planet-planet collision, planet-star collision, or planet ejection.  Considering the first planet loss only (so that the total number of planets is reduced from three to two), we introduce the variables $f_{\rm pp}$, $f_{\rm ps}$, $f_{\rm ej}$, which quantify the fractions of three-planet systems resulting in planet-planet collisions, planet-star collisions, and planet ejections respectively.  Fig.~\ref{fig:bratio} shows that the most common scattering outcome of the original three-planet systems is a planet-planet collision ($f_{\rm pp} \simeq 65 \%$), followed by planet ejection ($f_{\rm ej} \simeq 29 \%$) and planet-star collision ($f_{\rm ps} \simeq 6 \%$).  See also Table \ref{table1}.  Note that following the first planet loss, some systems may later lose a second planet, which is not included in the calculation of $f_{\rm pp}$, $f_{\rm ps}$, $f_{\rm ej}$.

In our numerical experiments, the two-planet systems arise entirely due to planet-planet collisions. The one-planet systems arise from a combination of ejections and collisions, with a modest preference for ejections for the first planet loss, and an overwhelming preference for ejections for the second planet loss.  Branching ratios separated into results for one and two-planet systems are also listed in Tables \ref{table_2p} and \ref{table_1p}.

\begin{table*}
 \centering
 \begin{minipage}{180mm}
  \caption{Parameters and results for the different sets of simulations. In all simulations, the initial innermost planet semi-major axis is sampled uniformly in $a_1 = [0.1-1]$AU, and the initial inclinations are sampled uniformly in the range $[0.1^\circ - 2^\circ]$.  For \fiducial, \eqmass, \lognorm, and \fidkthree, the initial eccentricities are sampled in the range $[0.01-0.05]$. For \fidkfive, we increase the initial eccentricities so that instabilities occur within a practical amount of time (see Section \ref{sec:K}).  The \lognorm\ simulations have all masses sampled from a log-normal distribution, with a mean of $\log (m_{\rm p} /  \mjup) = 0$, width $0.5$, and upper and lower limits of $m_{\rm p} = 4 \mjup$ and $m_{ \rm p} = 0.25 \mjup$.  The columns in the table are (from left to right), the simulation name, number of simulations ($N_{\rm run}$), choice of planet masses, initial orbital spacing in units of mutual Hill radii ($K$, see equations [\ref{eq:aspace}]-[\ref{eq:Rmut}]), fraction of one and two-planet systems produced at the end of ``Phase 2'' of the integration ($f_{\rm 1p}$ and $f_{\rm 2p}$).  The columns labeled $f_{\rm pp}$, $f_{\rm ps}$, and $f_{\rm ej}$ indicate the fraction of systems in which the first (or only) planet loss resulted from a planet-planet collision, planet-star collision, or planet ejection respectively (see Fig.~\ref{fig:bratio}).}
  \begin{tabular}{@{}lllllllllllll@{}}
  \hline
  \hline
Name & $N_{\rm run}$ & Initial Masses $(\mjup)$ & $K$  & $f_{\rm 1p}$ & $f_{\rm 2p}$ & $f_{\rm pp}$ & $f_{\rm ps}$ & $f_{\rm ej}$ \\
\hline
\fiducial & 3313 & $0.5,1.0,2.0$ & 4 & 0.52 & 0.48 & 0.65 & 0.06 & 0.29  \\ 
\eqmass & 973 & $0.9,1.0,1.1$ & 4 & 0.34 & 0.66 & 0.78 & 0.10 & 0.12   \\ 
\lognorm & 965 & See caption & 4 & 0.50 & 0.49 & 0.69 & 0.00 & 0.31  \\ 
\fidkthree & 986 & $0.5,1.0,2.0$ & 3 & 0.60 & 0.41 & 0.76 & 0.00 & 0.24  \\ 
\fidkfive & 949 & $0.5,1.0,2.0$ & 5 & 0.50 & 0.48 & 0.63 & 0.00 & 0.37  \\ 
\hline
\hline
\label{table1}
\end{tabular}
\end{minipage}
\end{table*}

\begin{table*}
 \centering
 \caption{Scattering outcomes and properties of the two-planet systems at the end of ``Phase 2'' of the integration.  The quantities $f_{\rm pp,2p}$, $f_{\rm ps,2p}$, $f_{\rm ej,2p}$ are the branching ratios for planet-planet collisions, planet-star collisions, and planet ejections respectively (see Fig.~\ref{fig:bratio}).  The remaining columns show the inner and outer planet eccentricities ($\ein, \eout$), the mutual inclination $\Imut$, and the semi-major axis ratio $\alpha \equiv \aout / \ain$.  The superscript ``ave'' denotes the unweighted average, and the superscripts $50, 90$ indicate the 50th (median), and 90th percentiles.}
 
  \begin{tabular}{@{}llllllllllllllll@{}}
 
  \hline
  \hline
Name & $f_{\rm pp,2p}$ & $f_{\rm ps,2p}$ & $f_{\rm ej,2p}$ & $\ein^{\rm ave}$ & $\ein^{50}$ & $\ein^{90}$ & $\eout^{\rm ave}$ & $\eout^{50}$  & $\eout^{90}$  & $\Imut^{\rm ave}$ $(\mathrm \circ)$ & $\Imut^{50}$ $(\mathrm \circ)$  & $\Imut^{90}$ $(\mathrm \circ)$ & $\alpha_{\rm ave}$ & $\alpha_{50}$ & $\alpha_{90}$ \\
\hline
\fiducial  & 0.48 & 0.0 & 0.0 & 0.15 & 0.11 & 0.34 & 0.12 & 0.09 & 0.27 & 5.25 & 2.46 & 13.26 & 2.46 & 2.06 & 3.02 \\ 
\eqmass & 0.66 & 0.0 & 0.0 & 0.18 & 0.14 & 0.38 & 0.15 & 0.12 & 0.3 & 7.16 & 3.01 & 18.39 & 2.62 & 2.04 & 3.69 \\ 
\lognorm & 0.49 & 0.0 & 0.0 & 0.14 & 0.1 & 0.31 & 0.12 & 0.08 & 0.25 & 5.20 & 2.28 & 14.73 & 2.3 & 2.01 & 2.74 \\ 
\fidkthree & 0.40 & 0.0 & 0.0 & 0.13 & 0.08 & 0.32 & 0.09 & 0.06 & 0.22 & 4.99 & 2.34 & 12.56 & 1.93 & 1.64 & 2.5 \\ 
\fidkfive & 0.48 & 0.0 & 0.0 & 0.15 & 0.12 & 0.31 & 0.15 & 0.11 & 0.33 & 5.14 & 2.16 & 12.02 & 2.63 & 2.37 & 3.42 \\ 
\hline
\hline
\label{table_2p}

\end{tabular}
\end{table*}

\begin{table*}
 \centering
 \begin{minipage}{180mm}
  \caption{Scattering outcomes and properties of the one-planet systems at the end of ``Phase 2'' of the integration.  The quantities $\fppfirst$, $\fpsfirst $ and $\fejfirst$ denote the branching ratios for the first planet loss (see Fig.~\ref{fig:bratio}). The quantities $\fppsec$, $\fpssec$ and $\fejsec$ have identical meanings, but for the second planet loss.  The remaining columns indicate the average, median, 10th, and 90th percentile eccentricities.}
  \begin{tabular}{@{}lllllllllll@{}}
  \hline
  \hline
Name & $\fppfirst$ & $\fpsfirst $ & $\fejfirst $ & $\fppsec$ & $\fpssec$ & $\fejsec$ & $e_{\rm ave}$ & $e_{50}$ & $e_{10}$ & $e_{90}$ \\
\hline
\fiducial & 0.17 & 0.06 & 0.29 & 0.01 & 0.03 & 0.48 & 0.30 & 0.27 & 0.10 & 0.55 \\ 
\eqmass & 0.12 & 0.10 & 0.11 & 0.01 & 0.02 & 0.31 & 0.47 & 0.42 & 0.23 & 0.78 \\ 
\lognorm & 0.2 & 0.0 & 0.3 & 0.04 & 0.0 & 0.46 & 0.30 & 0.26 & 0.10 & 0.56 \\ 
\fidkthree & 0.35 & 0.0 & 0.24 & 0.06 & 0.0 & 0.53 & 0.29 & 0.26 & 0.10 & 0.55 \\ 
\fidkfive & 0.15 & 0.0 & 0.36 & 0.03 & 0.0 & 0.47 & 0.30 & 0.26 & 0.09 & 0.55 \\
\hline
\hline
\label{table_1p}
\end{tabular}
\end{minipage}
\end{table*}

\begin{table*}
 \centering
 \begin{minipage}{180mm}
  \caption{Properties of both one and two-planet systems produced from the \fourp\ simulations (see Section \ref{sec:fourp}). The columns labeled $f_{\rm 1p}$ and $f_{\rm 2p}$ have the same meanings as in Table \ref{table1}.  The next three columns indicate the average, 50th and 90th percentile  eccentricities of the one-planet systems. The remaining columns indicate properties of two-planet systems, with the same notation as in Table \ref{table_2p}.}
  \begin{tabular}{@{}llllllllllllllllll@{}}
  \hline
  \hline
Name & $f_{\rm 1p}$ & $f_{\rm 2p}$ & $e_{\rm ave}$ & $e_{50}$ & $e_{90}$ & $e_{\rm in}^{\rm ave}$ & $e_{\rm in}^{50}$ & $e_{\rm in}^{90}$ & $e_{\rm out}^{\rm ave}$ & $e_{\rm out}^{50}$ & $e_{\rm out}^{90}$ & $I_{\rm mut}^{\rm ave}$ & $I_{\rm mut}^{50}$ & $I_{\rm mut}^{90}$ & $\alpha_{\rm out}^{\rm ave}$ & $\alpha_{\rm out}^{50}$ & $\alpha_{\rm out}^{90}$  \\
\hline
\fourp & 0.75 & 0.23& 0.35 & 0.31 & 0.64 & 0.2 & 0.17 & 0.4 & 0.2 & 0.17 &  0.4 & 7.21 & 4.01 & 15.76 & 4.13 & 2.87 & 4.98 \\ 
\hline
\hline
\label{table_4p}
\end{tabular}
\end{minipage}
\end{table*}

\subsection{Properties of One and Two-Planet Systems} \label{sec:parameters}

In this section we analyze the orbital characteristics of the one and two-planet systems at the end of Phase 2 of the integration.  In most figures we also show properties of observed WJ systems; however, we delay comparison with the observed systems until Section \ref{sec:obs}.

\begin{figure*}
\centering 
\includegraphics[width=\textwidth]{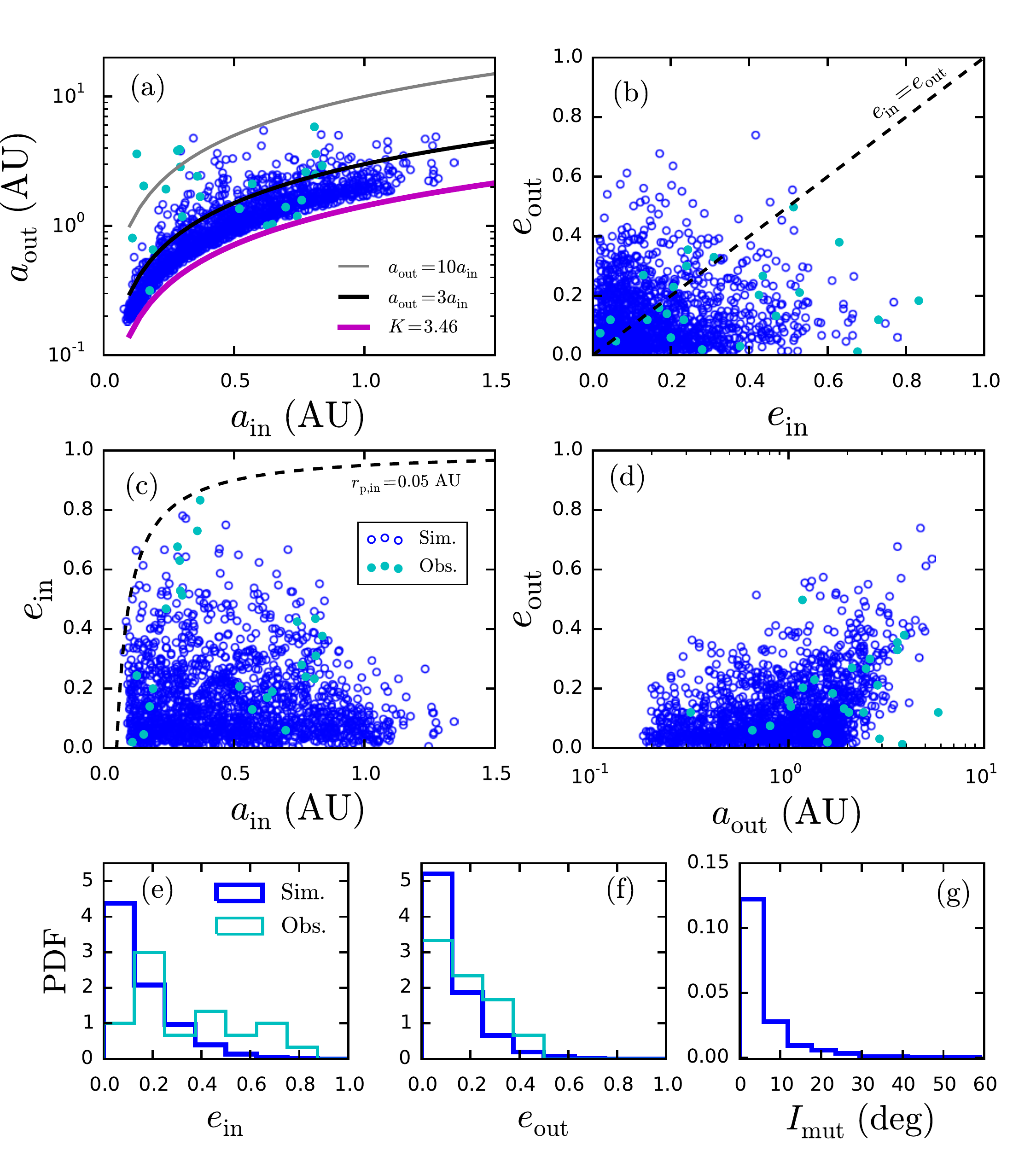}
\caption{Two-planet systems (open blue circles) that emerge from three-planet scattering (from the \fiducial\ simulations), along with the 24 observed WJ systems with external giant planet companions (filled cyan circles).  {\it Panel (a):} $\ain$ versus $\aout$. Scattering tends to result in compact two-planet systems. Over $90 \%$ of systems satisfy $\aout/\ain \lesssim 3$. {\it Panel (b):} $\ein$ versus $\eout$.  Scattering results in a wide range of eccentricities for both planets, with a slight preference for a higher eccentricity of the inner planet.  {\it Panel (c):} $\ain$ versus $\ein$.  Also plotted is a contour of constant pericenter distance $r_p = 0.05$ AU; systems below this curve are not expected to undergo inward migration due to tides raised on the planet.  {\it Panel (d):} $\aout$ versus $\eout$.  {\it Panels (e)-(f):} Histograms of eccentricities of the inner and outer planets $\ein$, $\eout$.   {\it Panel (g):} Mutual inclination between the inner and outer planet $I_{\rm mut}$.  Scattering results in low inclinations, with $90 \%$ of systems satisfying $I_{\rm mut} \lesssim 13^\circ$.}
\label{fig:2pscatter}
\end{figure*}

Figure \ref{fig:2pscatter} shows the properties of the two-planet systems.  The subscript ``in'' indicates the inner planet and ``out'' indicates the outer planet.  We comment on the key features here, but see the caption and Table \ref{table_2p} for additional details.  In Fig.~\ref{fig:2pscatter}a, we present the relative spacing, $\aout$ versus $\ain$.  The two-planet systems are quite closely-spaced, with $90 \%$ of systems satisfying $\aout / \ain < 3$.  In Fig.~\ref{fig:2pscatter}c, we show semi-major axis versus eccentricity for the inner planet ($\ain$ versus $\ein$).  Smaller values of $\ain$ lead to a wider range of eccentricities.  The eccentricity distribution in Fig.~\ref{fig:2pscatter}e is peaked at small values, and has mean and 90th percentile values $e_{\rm in}^{\rm ave} = 0.15$, $e_{\rm in}^{90} = 0.34$.  The mutual inclinations of the two-planet systems (Fig.~\ref{fig:2pscatter}g) have mean and 90th percentiles $I_{\rm mut}^{\rm ave} = 5.25^\circ$ and $I_{\rm mut}^{90} = 13.26^\circ$. The typically low eccentricities and inclinations are consistent with the fact that the two-planet systems arose entirely due to planet-planet collisions \citep{ford2001}.

\begin{figure*}
\centering 
\includegraphics[width=\textwidth]{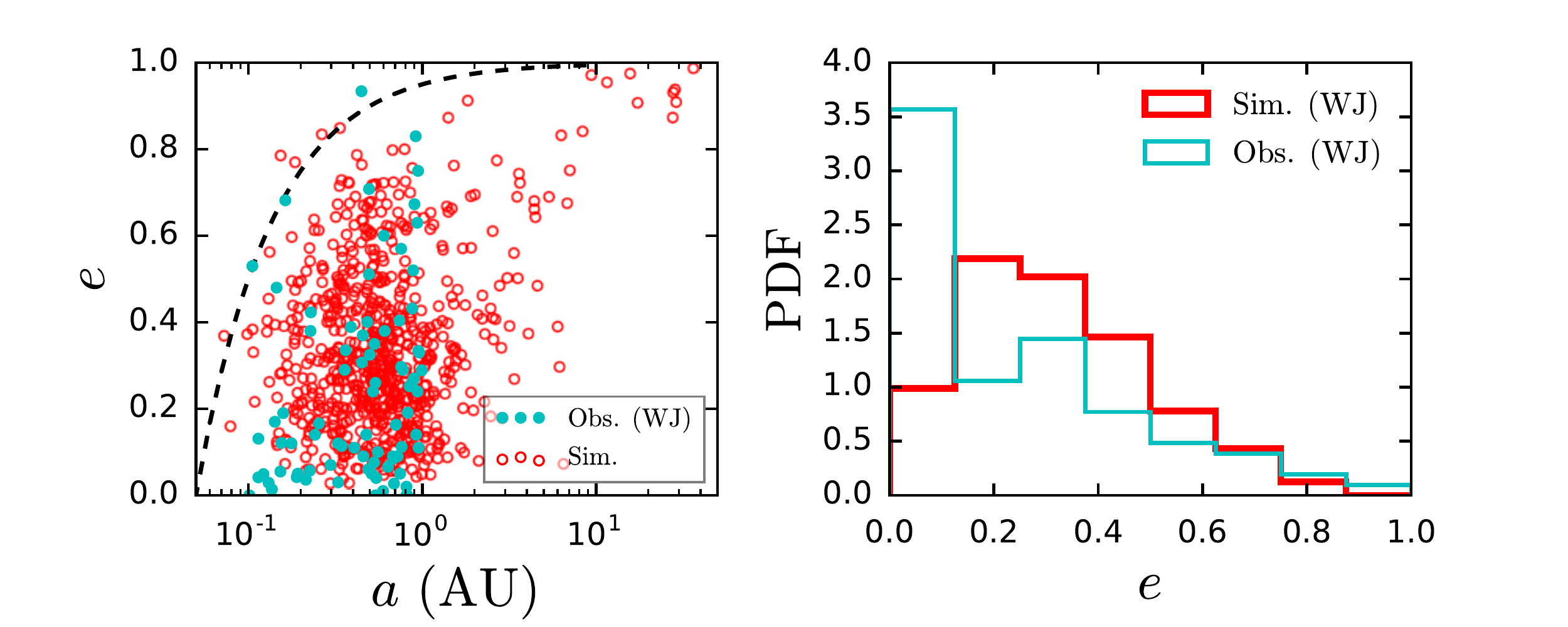}
\caption{One-planet systems that are formed from the \fiducial\ simulations (open red circles), along with observed ``solitary WJs'' (without any identified giant planet companions), shown as filled cyan circles.  This observational sample consists of 83 systems at present. {\it Left}: Eccentricity versus semi-major axis, along with a curve of constant pericenter $r_p = 0.05$ AU for reference.  Systems below this dashed curve are not expected to undergo migration due to tides raised on the planet.  {\it Right:} Histograms of WJ eccentricities (only systems within $1$ AU are shown). The simulated eccentricity distribution peaks near $e \simeq 0.2$ with a tail extending to $\sim 0.9$.}
\label{fig:1p}
\end{figure*}

Figure \ref{fig:1p} shows the properties of the one-planet systems.   Most of these planets are WJs, but a small fraction have semi-major axes greater than 1 AU.  Compared to the two-planet systems, the degree of eccentricity excitation for the one-planet systems is higher, due to the fact that each one-planet system has previously suffered at least one ejection.  The distribution of eccentricities shown in Fig.~\ref{fig:1p} peaks near $0.25$ and has a long tail.  The mean and 90th percentile eccentricities of the single-planet systems are $e_{\rm ave} = 0.3$ and $e_{90} = 0.55$ respectively.

\subsection{Dependence on initial spacing $K$}
\label{sec:K}
We now examine how the results depend on the initial spacing parameter $K$.  Previous work has shown that when planet collisions are infrequent, $K$ primarily affects the time for instabilities \citep{chambers1996}, rather than the degree of eccentricity excitation itself.  However, the situation is unclear when planet collisions are common.

The \fiducial\ simulations have $K = 4$.  To explore how the results depend on $K$, we perform additional sets of $\sim 1000$ simulations with $K = 3$ and $K = 5$ (titled \fidkthree\ and \fidkfive\ respectively), keeping the other parameters the same as \fiducial.  However, for \fidkfive, the timescale for instabilities to develop is sufficiently long to be impractical, so we slightly increase the initial eccentricities of the planets: We set the initial  $e_{1} = 0.08$ and assign the $e_2$ and $e_3$ so that $a_3 (1 - e_3) - a_2(1 + e_2) = 3 R_{\rm H,mut}$.  Depending on the mass ordering, the initial eccentricities of the outer planets are at most $\sim 0.11$ Such eccentricities may conceivably develop due to planet-disk interactions.

Inspection of Tables \ref{table1} - \ref{table_1p} reveals that the value of $K$ does not drastically affect the results.  As expected, the fraction of planet-planet collisions is highest for the smallest $K$, with $f_{\rm pp} = 0.76$ for $K = 3$, compared to $f_{\rm pp} = 0.63$ for $K = 5$. The eccentricities listed in Tables \ref{table_2p} and \ref{table_1p} do not differ appreciably. The final spacing of the two-planet systems ($\alpha \equiv \aout / \ain$) is mildly dependent on $K$, with average $\alpha_{\rm ave} = 1.93, 2.46, 2.63$ for $K = 3,4,5$, and 90th percentile spacing $\alpha_{90} = 2.5, 3.02, 3.42$.

\subsection{Dependence on initial innermost semi-major axis} \label{sec:a1}
Next we discuss how the scattering results depend on the initial innermost planet semi-major axis $a_1$. Taking \fiducial\ (with $K = 4$) we split the simulations ($3313$ total) into four bins of $a_1$, each with width $0.225$ AU and centered at $0.21$, $0.44$, $0.66$, and $0.89$ AU.

\begin{figure}
\centering 
\includegraphics[scale=0.7]{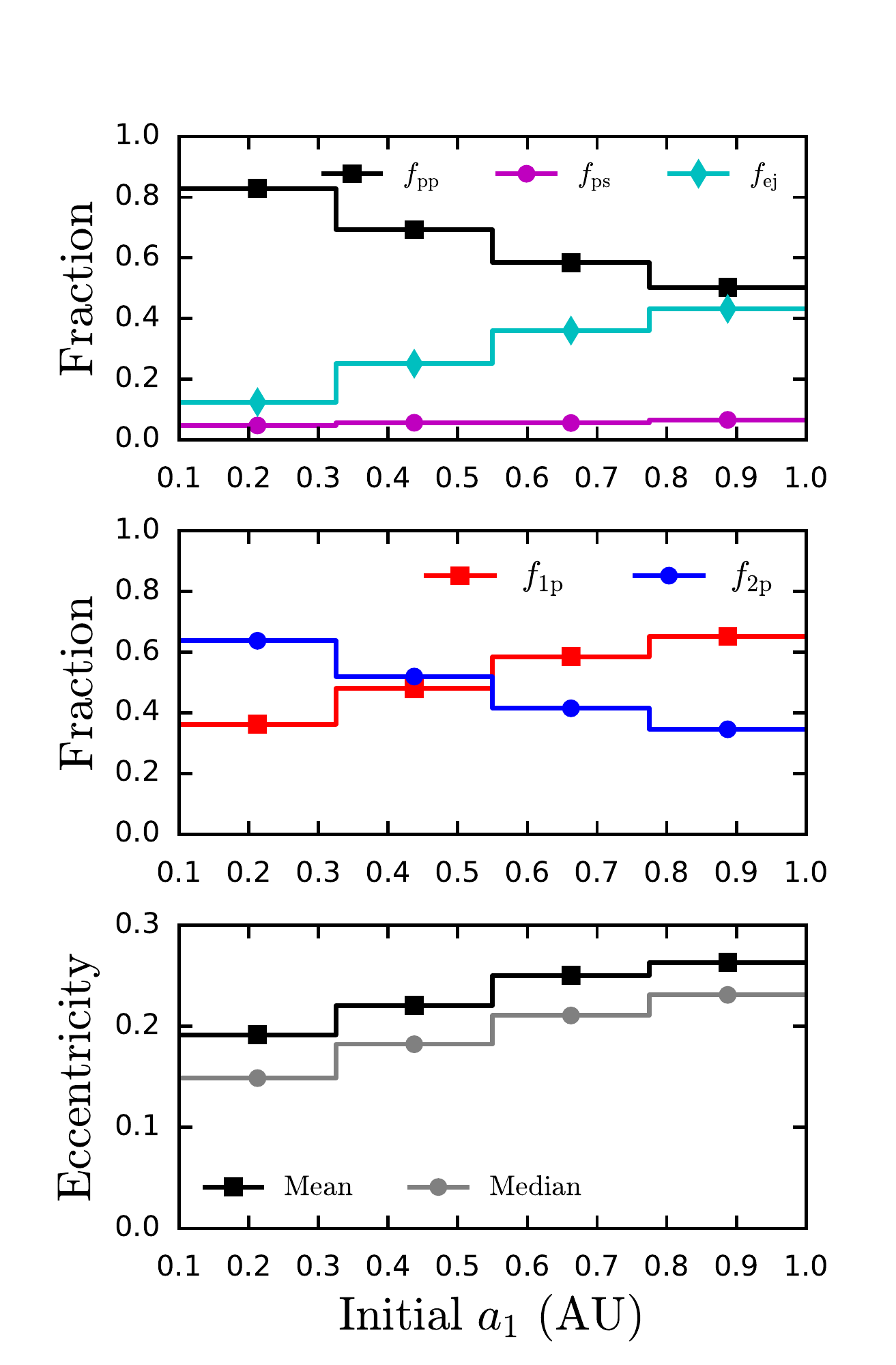}
\caption{Dependence of the scattering outcomes on the initial inner planet semi-major axis $a_1$, from the \fiducial\ simulations.   {\it Top panel:} Fractions of systems resulting in planet-planet collisions ($f_{\rm pp}$), planet-star collisions ($f_{\rm ps}$), and planet ejections ($f_{\rm ej}$) for the first (or only) planet loss (see Fig.~\ref{fig:bratio}).  Planet-planet collisions dominate at small values of $a_1$.  As $a_1$ increases, the frequency of planet ejections increases. {\it Middle panel:} Fractions of one-planet ($f_{1 \rm p}$) and two-planet ($f_{2 \rm p}$) systems produced via scattering.  Due to the increasing frequency of ejections with $a_1$, the fraction of one-planet systems increases with $a_1$.  {\it Bottom panel:} Median and mean eccentricities of both one and two-planet systems. Eccentricities increase with $a_1$, due to the increasing fraction of one-planet systems. }
\label{fig:bratio_ecc_binned}
\end{figure}

Figure \ref{fig:bratio_ecc_binned} shows how the results depend on $a_1$.  The top panel shows the frequencies of collisions and ejections.  In the first bin, centered at $0.21$ AU, planet-planet collisions are very common, with $f_{\rm pp} \simeq 82 \%$ and $f_{\rm ej} \simeq 15 \%$.  As $a_1$ increases, the frequency of ejections increases, so that in the last bin, centered at $0.89$ AU, $f_{\rm pp} \simeq 50 \%$. The fraction of planet-star collisions remains nearly constant at $f_{\rm ps} \simeq 6 \%$ across all bins.

The dependency of $f_{\rm pp}$ and $f_{\rm ej}$ on $a_1$ leads to a variation of the relative numbers of one and two-planet systems, as shown in the middle panel of Fig.~\ref{fig:bratio_ecc_binned}.  In the first $a_1$ bin, we have $f_{\rm 1p} = 37 \%$ and $f_{\rm 2p} = 63 \%$.  The larger value of $f_{\rm 2p}$ at small $a_1$ is a direct consequence of the fact that collisions are most common close to the host star, leading to systems with lower eccentricities, and hence, more stable two-planet systems.  As $a_1$ increases, the number of ejections increases, so that $f_{\rm 1p}$ increases, reaching $f_{\rm 1p} = 63 \%$ and $f_{\rm 2p} = 37 \%$ in the last bin.  This dependency of collisions/ejections with $a_1$ leads to a slight increase in the average eccentricity with $a_1$, as shown in the bottom panel of Fig.~\ref{fig:bratio_ecc_binned}.   Note that the eccentricities shown in Fig.~\ref{fig:bratio_ecc_binned} include both the one and two-planet systems.

\begin{figure*}
\centering 
\includegraphics[width=\textwidth]{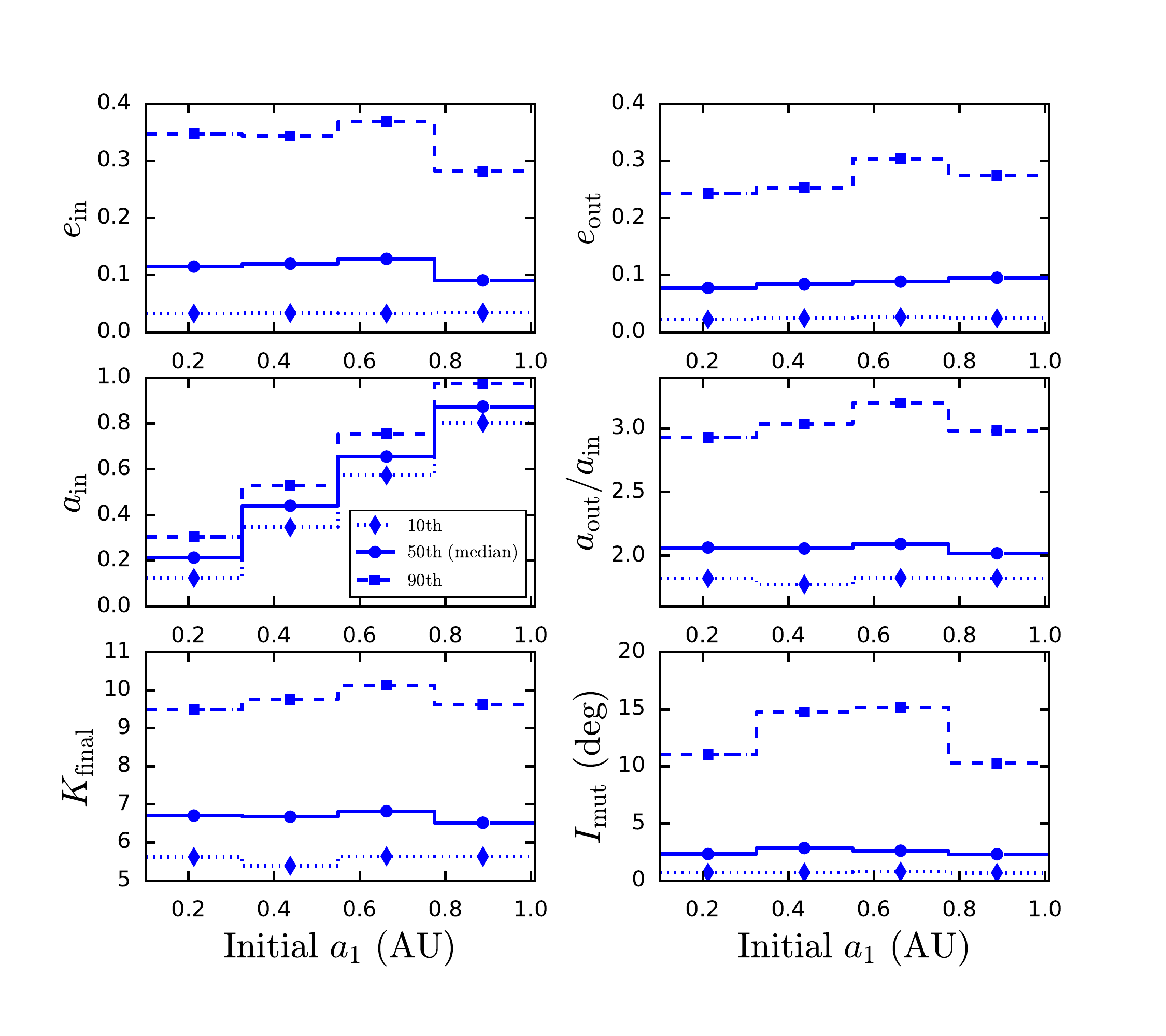}
\caption{Dependence of two-planet system properties on $a_1$ of the initial three-planet system, from the \fiducial\ simulations.  The vertical axes show the 10th, 50th, and 90th percentiles of various quantities, as labeled. {\it Top left and right:} Inner and outer planet eccentricities. The initial value of $a_1$ does not strongly affect the $\ein$ and $\eout$ statistics, with a median $\ein$ of $\sim 0.1$ across all bins, and slightly lower for $\eout$.  {\it Middle left:} Inner planet semi-major axis $\ain$ of the surviving two-planet systems.  Scattering typically results in $\ain$ within a factor of $\sim 2$ of the initial inner semi-major axis.  {\it Middle right:} $\aout/\ain$. Scattering tends to produce closely-spaced two-planet systems ($\aout / \ain \simeq 2-3$), with more hierarchical systems rare. {\it Bottom left:} The spacing of the final two-planet systems in units of mutual Hill radii, with a median value $K_{\rm final} \simeq 6.5$ across all bins. {\it Bottom right:} Mutual inclinations of the two-planet systems.  Scattering typically results in low inclinations, with a median of $\sim 2^\circ$ across all bins. }
\label{fig:2p_binned}
\end{figure*}

Figure \ref{fig:2p_binned} illustrates how the initial $a_1$ affects the properties of the final two-planet systems.  The initial $a_1$ primarily determines the final semi-major axis, since scattering typically limits changes in semi-major axis to factors of order unity. The final planet eccentricities, mutual inclination, and relative spacing depend weakly on $a_1$.  A  similar weak dependence on $a_1$ for one-planet systems is also found (not shown).

\subsection{Dependence on Planet Masses} \label{sec:masses}
Finally, we explore how the scattering results depend on planet masses.   When planet masses are unequal, scattering leads to fewer close encounters between planets, since a massive planet can easily eject a low-mass planet.  For equal-mass planets, ejection becomes more difficult and planets may suffer a higher number of close encounters, leading to a more efficient eccentricity excitation \citep{ford2008}.  Besides \fiducial\ (with masses $0.5, 1, 2 \mjup$), we conduct two additional sets of simulations, with all parameters (except for the masses) chosen/sampled identically to \fiducial. We consider the case of nearly equal mass planets, with $m_{\rm p} = 0.9, 1, 1.1 \mjup$, and refer to this set of simulations as \eqmass.  We also consider masses sampled from a log-normal distribution, referred to as \lognorm, with mean $1 \mjup$, width $\log_{10} (m_{\rm p}/\mjup) = 0.5$, and upper and lower limits $0.25 \mjup$ and $4 \mjup$.  These choices are somewhat arbitrary, but qualitatively mimic the observed distribution of giant planet masses.  In all simulations, the ordering of the three masses is randomly assigned.

\begin{figure*}
\centering 
\includegraphics[width=\textwidth]{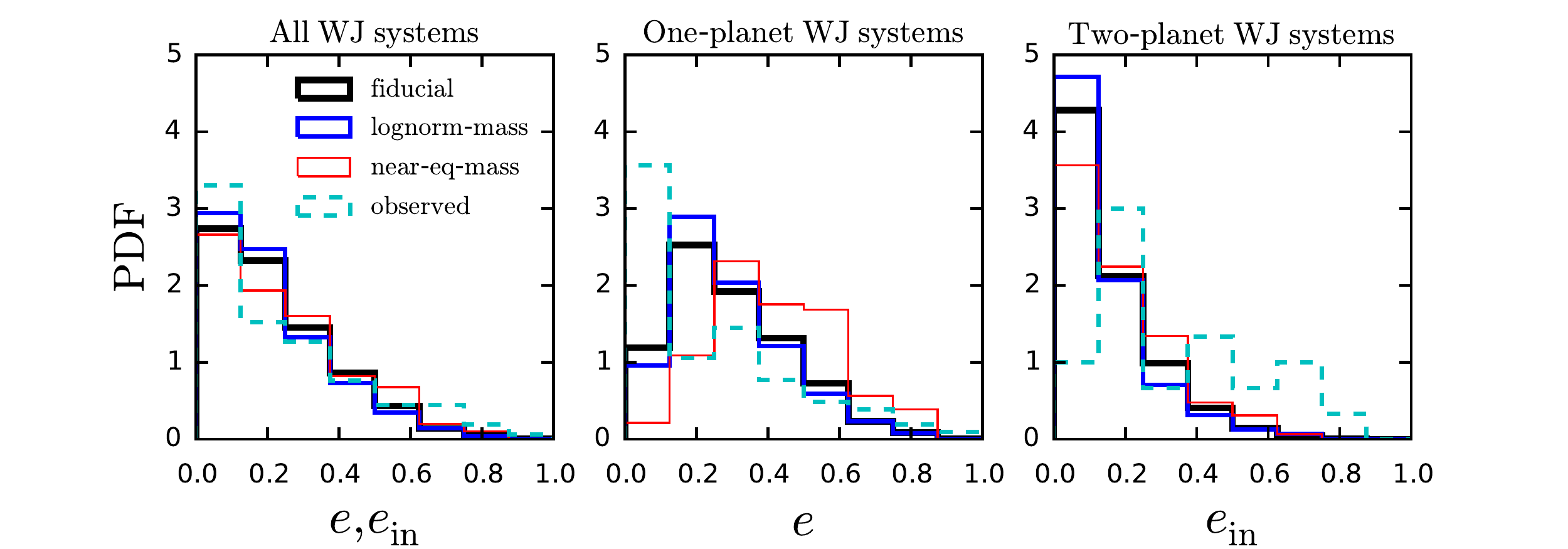}
\caption{Eccentricities of WJs after scattering, illustrating the dependence on planet masses.  The middle and right panels show the eccentricities of one-planet systems and the inner planet eccentricity of the two-planet systems respectively (similarly depicted in Figs.~\ref{fig:2pscatter} and \ref{fig:1p} for the \fiducial\ simulations). The left panel combines the eccentricity distributions shown in the middle and right panels.  Black histograms indicate the \fiducial\ run, while blue and red indicate \lognorm\ and \eqmass\ respectively.  Observed WJs are shown as the dashed cyan histograms. The \eqmass\ simulations exhibit the largest degree of eccentricity excitation.}
\label{fig:hist_masses}
\end{figure*}

Figure \ref{fig:hist_masses} compares the eccentricity distributions of \fiducial, \eqmass, and \lognorm, showing the combined eccentricity distribution of one and two-planet systems (left panel), as well as eccentricities split into one-planet and two-planet systems (middle and right panels). We only show results for simulated one-planet systems with $a < 1$AU in order to compare with observed WJ systems in Section \ref{sec:obs}.  The results for the two planet systems show the inner planet eccentricity only.  As expected, \eqmass\ produces the highest eccentricities (see also Tables \ref{table_2p} - \ref{table_1p}).  Only minor differences are observed between \fiducial\ and \lognorm.
 
The choice of planet masses affects the relative spacing of the two-planet systems, but only marginally.  Inspecting Table \ref{table_2p}, the average values of $\aout / \ain$ for \fiducial, \eqmass, and \lognorm\ are $2.46$, $2.62$, and $2.3$ respectively.  As a result, the finding that in-situ scattering produces very compact two-planet systems is robust with respect to the choice of planet masses.

\subsection{Scattering of Four Planets}\label{sec:fourp}
All scattering experiments discussed thus far began with three unstable planets.  We also constructed a set of four-planet systems,  with masses $m_{\rm p} = (0.5,1,1.25,2) \mjup$ (placed in random order), and all other parameters chosen the same as \fiducial.  We term this set of simulations \fourp, integrate the systems through ``Phase 1'' and ``Phase 2,'' and analyze the properties of the surviving planets.  All four-planet systems are destabilized, and the fractions of one, two, and three-planet systems at the end of ``Phase 2'' are $0.75$, $0.23$, and $0.02$ respectively (see also Table \ref{table_4p}). Adding an additional planet allows more possibilities for ejections. As a result, while \fiducial\ (with three planets) produces roughly equal numbers of one and two-planet systems, \fourp\ produces over three times more one-planet systems compared to two-planet systems.  The average eccentricity for one-planet systems from \fourp\ is $e_{\rm ave} = 0.35$ (see also Table \ref{table_4p}), compared with $e_{\rm ave} = 0.3$ from \fiducial. 

\begin{figure}
\centering 
\includegraphics[scale=0.65]{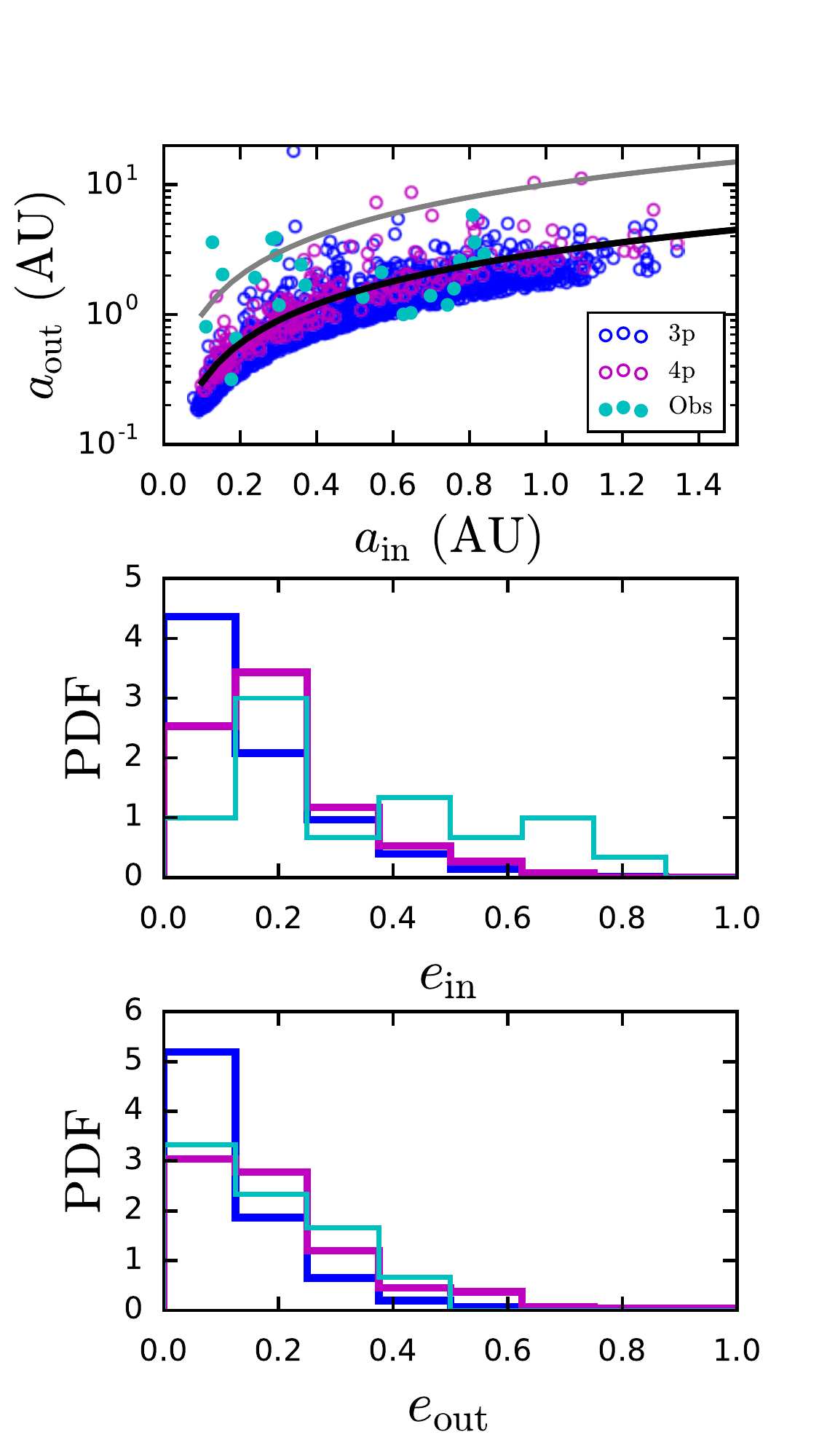}
\caption{Properties of two-planet systems produced from \fourp\ (labeled ``4p''), along with those from \fiducial\ for reference (labeled ``3p'').  Observed two-planet systems are shown in cyan  (see also Section \ref{sec:obs}). \fourp\ consists of systems of four initially unstable planets with masses $0.5,1,1.25,2 \mjup$, and all other parameters identical to \fiducial.  {\it Top:} Inner versus outer planet semi-major axis.  The black and grey curves indicate $\aout / \ain = 3$ and 10. {\it Middle and bottom:} Eccentricities of the inner and outer planets, with the same color scheme as the top panel. \fourp\ produces more eccentric planets and wider spacing compared to \fiducial.}
\label{fig:3p_4p}
\end{figure}

Figure \ref{fig:3p_4p} shows the two-planet systems produced by \fourp, along with the \fiducial\ results for reference.  The top panel of Fig.~\ref{fig:3p_4p} shows that \fourp\ produces more hierarchical systems, with mean value of $\alpha = \aout / \ain = 4.13$.  The remaining panels of Fig.~\ref{fig:3p_4p} show the inner and outer planet eccentricities.  Unsurprisingly, the eccentricities of both planets tend to be higher, with an average inner planet eccentricity $e_{\rm in}^{\rm ave} \simeq 0.2$ for \fourp\ compared to $e_{\rm in}^{\rm ave} \simeq 0.15$ for \fiducial.

\section{Comparison with Observations}\label{sec:obs}
The Exoplanet Orbit Database lists 106 RV giant planet systems ($m_{\rm p} \sin i > 0.3 \mjup$) with semi-major axes in the range $[0.1 {\rm AU } - 1]{\rm AU}$, and with eccentricity and mass measurements listed.  Of these 106 systems, 23 have an external giant companion with a characterized orbit and minimum mass.  The majority of these two-planet systems were previously discussed by \cite{antonini2016}, in the context of high-eccentricity migration, and by \cite{anderson2017}, in the context of secular perturbations.  We augment this sample with the well-known Kepler-419 system \citep{dawson2014kep419}, which is not flagged by our search criteria due to lack of planet mass listings on exoplanets.org. This yields a sample of 24 WJs with an external giant planet companion.

In the following we compare our scattering results to this observed sample of WJs.  Unless otherwise stated, we will refer to our three-planet scattering results. 

\subsection{Eccentricities}\label{sec:ecc}
As discussed above, in-situ scattering of three planets leads to a range of eccentricities in the final (post-scattering) planetary orbit(s).   The left panel of Fig.~\ref{fig:hist_masses} presents the combined eccentricity distributions of the one and two-planet systems, consisting of $e$, the eccentricities of single WJs, and $\ein$, the inner planet eccentricities of the two-planet systems.  We see that in-situ scattering of three planets with a variety of mass choices reproduces the observed distribution fairly well.

However, separating the one and two-planet systems tells a somewhat different story.  Recall that the two-planet systems tend to have relatively low eccentricities due to the fact that they formed exclusively via planet-planet collisions (see Table \ref{table_2p}).  Comparing the observed and simulated eccentricity distributions in Figs.~\ref{fig:2pscatter}e and \ref{fig:hist_masses}, we see that although scattering is able to produce the full range of observed eccentricities, there is a clear discrepancy in the shapes of the distributions: the observed distribution is much flatter.  This suggests that in-situ scattering of three planets cannot adequately account for the full sample of observed two-planet systems.  The addition of a fourth planet alleviates, but does not completely resolve this problem (see the middle panel of Fig.~\ref{fig:3p_4p}).

The single-planet systems, which suffered at least one ejection, tend to have much higher eccentricities.  Figs.~\ref{fig:1p} and \ref{fig:hist_masses} reveal that the in-situ scattering of three planets can reproduce the large eccentricities of the observed solitary WJs. But the observed peak at low eccentricities is not reproduced.

\begin{figure}
\centering 
\includegraphics[scale=0.45]{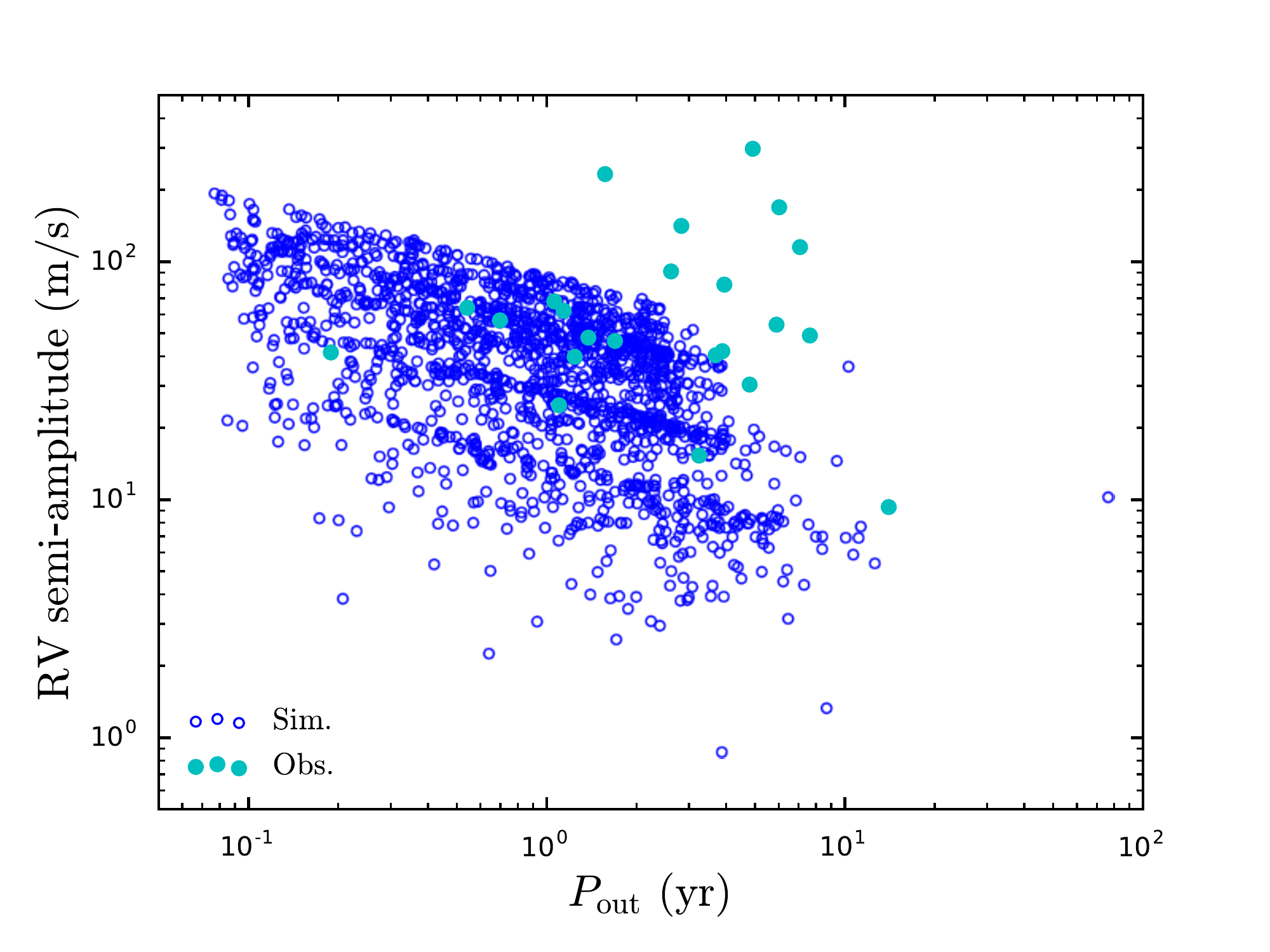}
\caption{Radial velocity semi-amplitude versus orbital period for the outer planet.  The \fiducial\ simulations are shown as open blue circles, while observed systems are shown as filled cyan circles.  An isotropic distribution of sky-projected inclinations has been assumed in calculating the expected radial velocity semi-amplitude.  }
\label{fig:rvamp}
\end{figure}

The tension between the observed and simulated eccentricity distributions of both the one and two planet systems may be relieved if some of the observed solitary WJs have an undetected  companion.  If some of the observed one-planet systems are actually two-planet systems, the number of circular one-planet systems may decrease, and the number of circular two-planet systems may increase, leading to a better match with the scattering simulations. To evaluate this possibility, we calculate the radial-velocity semi-amplitude of the outer planet in our simulated two-planet systems from \fiducial, given by \citep{cumming2008}
\be
K_{\rm RV} = \frac{28.4 \ {\rm m/s}}{\sqrt{1 - e^2}} \left(\frac{m_{\rm p} \sin i_{\rm sky}}{\mjup} \right)  \left(\frac{P}{1 \ {\rm yr}} \right)^{-1/3}  \left(\frac{M_\star}{\msun} \right)^{-2/3}.
\ee
In Fig.~\ref{fig:rvamp} we show the RV semi-amplitude versus orbital period of the outer planet in the two-planet systems, where we have assumed an isotropic distribution of sky-projected inclinations $i_{\rm sky}$.  Scattering produces a population of outer planets with RV semi-amplitudes mostly in the range $\sim 10-100 {\rm m} / {\rm s}$.  Due to the compactness of the two-planet systems, many systems have large RV amplitudes and short-orbital periods (with $P_{\rm out} \lesssim 1$ yr) which should be readily detected, but are typically not observed. We note, however, the possibility of undetected companions due to low-cadence RV observations.

\begin{figure*}
\centering 
\includegraphics[width=\textwidth]{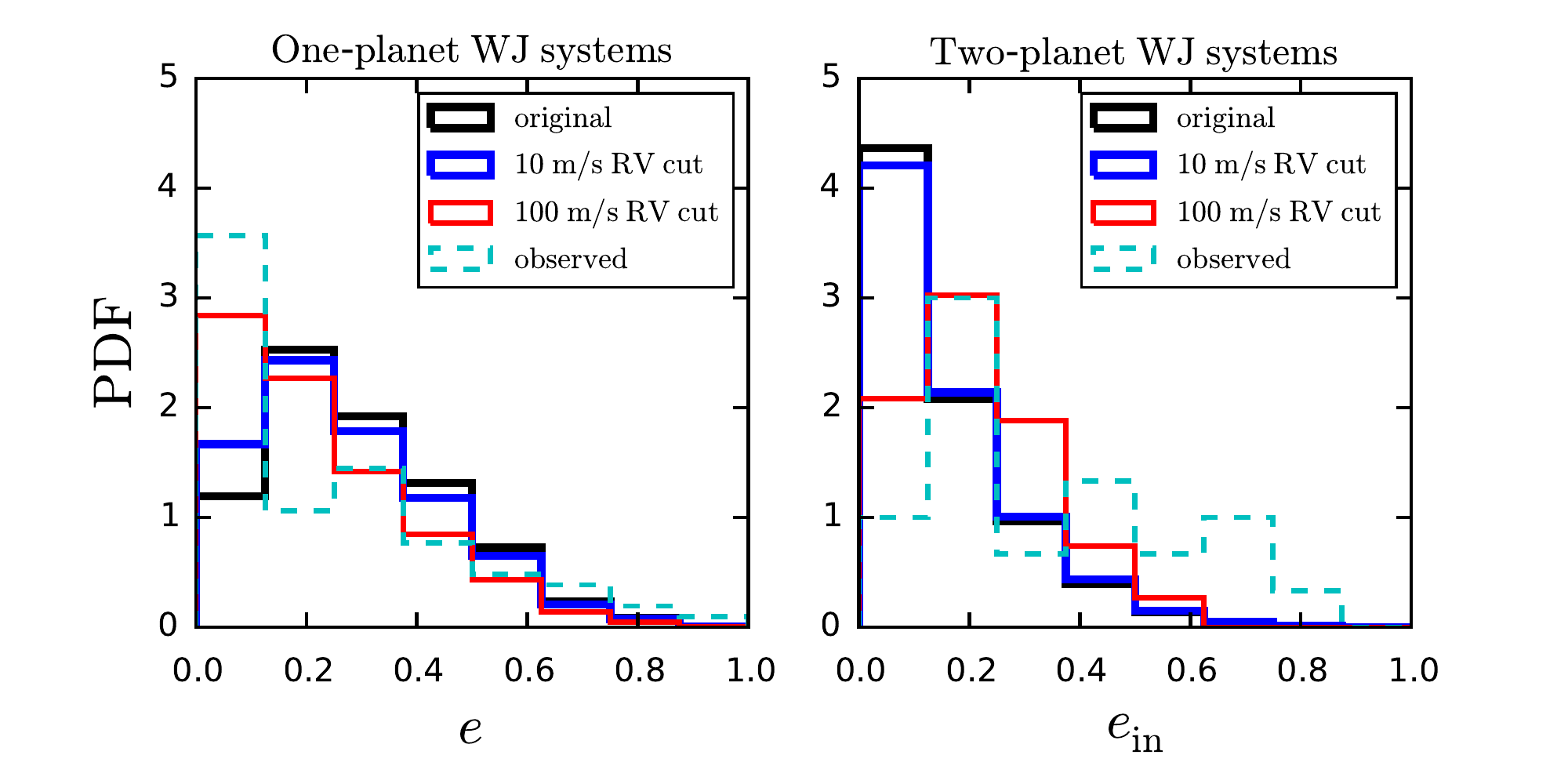}
\caption{The effect of imposing an RV cut on the \fiducial\ one-planet and two-planet systems.  The left panel depicts eccentricities of one-planet systems and the right panel depicts inner planet eccentricities of two-planet systems. The black histograms depict the original simulation results (no RV cut). We have calculated the expected RV semi-amplitudes of the outer planets, and classified any two-planet systems as one-planet systems if the outer planet has an RV semi-amplitude less than the specified cut.  Imposing a 10 m/s cut (blue histograms) only slightly affects the eccentricity distributions.  A 100 m/s cut (red histograms) yields better agreement between simulations and observations for both the one and two-planet systems, but planets with such large RV amplitudes are readily detected.}
\label{fig:ecc_rvcut}
\end{figure*}

To test whether undetected companions to solitary WJs may reconcile the disparities in eccentricity distributions, we apply RVs cuts of $10$ and $100$ m/s to the two-planet \fiducial\ simulations, and classify any two-planet systems as one-planet systems if the outer planet has an RV semi-amplitude less than the specified cut.  The results are shown in Fig.~\ref{fig:ecc_rvcut}.  A 10 m/s RV cut does not appreciably change the original eccentricity distributions. A 100 m/s RV cut does provide a significantly better agreement with observations.  However, since planets with tens of m/s RV amplitudes are readily discovered, there is unlikely to be a large population of undetected companions to observed single WJs with the properties obtained from our scattering simulations. Thus, we conclude that in-situ scattering is unable to separately reproduce the full observed eccentricity distributions of one and two-planet systems, in spite of the apparent agreement in the combined eccentricity distributions (left panel of Fig.~\ref{fig:hist_masses}).

Instead, the eccentricity distribution of one-planet systems is consistent with in-situ scattering having occurred in a significant fraction of systems, alongside a population of low-eccentricity planets that did not undergo scattering. We quantitatively explore this possibility in Section \ref{sec:stats}.

\subsection{Spacing and Mutual Inclinations of Two-Planet Systems}
As discussed above, scattering of three giant planets usually results in closely spaced systems, with $90 \%$ of the \fiducial\ two-planet systems satisfying $\aout / \ain < 3$.  Some observed systems are also quite closely spaced (see Fig.~\ref{fig:2pscatter}a), but there exists others that are much more hierarchical.  Out of the 24 observed two-planet systems, 7 have $\aout / \ain < 3$.

The mutual inclinations generated via scattering are usually low, with $90 \%$ of systems having inclinations less than $12^\circ-18^\circ$, and the median is $\sim 2^\circ - 3^\circ$.  With only a few exceptions, the observed giant exoplanet systems lack constraints on mutual inclinations at present.

\subsection{Relative Numbers of One and Two-Planet Systems} \label{sec:rel}
As mentioned previously, our observational sample consists of 83 WJs without a detected giant companion, and 24 WJs with a characterized external companion. This yields relative numbers of one and two-planet systems  $N_{1,\rm obs} / N_{2, \rm obs} \simeq 3.5$.  However, 8 of the 83 solitary WJs have a linear trend in the RV curve, indicating the possibility of an external companion.  If we assume the presence of a planetary companion in each of these 8 systems, then $N_{1,\rm obs} / N_{2, \rm obs} = 75 / 32 \simeq 2.3$.

The three-planet scattering experiments (i.e. \fiducial, \lognorm, \eqmass, \fidkthree, \fidkfive) yield relative numbers of single WJs and WJs with external companions $N_{1,\rm sim} / N_{2, \rm sim} \simeq 0.5 - 1.5$, where the range arises from the choice of planet masses and initial spacing.  As a result, in-situ scattering of three giant planets generally predicts an excess of WJs with a companion compared to observations.  Four-planet scattering (\fourp) yields $N_{1,\rm sim} / N_{2, \rm sim} \simeq 3.3$, in good agreement with observations.

It is useful to consider the extreme hypothesis that the majority of WJs form in systems of three giant planets, which then undergo scatterings.  In this scenario, there are three ways of reconciling the discrepancy between $N_{1,\rm obs} / N_{2, \rm obs}$ and $N_{1,\rm sim} / N_{2, \rm sim}$:
\begin{enumerate}
\item A fraction of the 2-planet systems generated through scattering are actually unstable.  As discussed in Section \ref{sec:scatoutcome}, $\sim 10 \%$ or more of the two-planet systems are expected to eventually undergo instabilities.

\item A fraction of the single WJs actually have an external companion, most likely having $\aout / \ain < 3$.  The possibility of undetected outer companions was already discussed in Section \ref{sec:ecc}.  

\item Finally, a third way to resolve the difference in multiplicity frequency is to modify the initial conditions.  For practical purposes we have restricted the initial spacings to initial spacings $K = 3-5$. Wider spacings may lead to different relative frequencies of collisions and ejections, thus affecting the relative numbers of one and two-planet systems. Such an exploration is beyond the scope of this work.
\end{enumerate}
  
Alternatively, agreement between the predicted and observed relative frequency of one and two-planet systems is not expected if in-situ-scattering occurred in only a fraction of WJ systems.  Indeed,  the inability to separately match the eccentricity distributions of one and two-planet systems (see Section \ref{sec:ecc}) probably implies in-situ scattering is not a universal outcome of planet formation. We discuss this possibility in the following section.

\subsection{Two-Population Mixture Model for Solitary WJ Eccentricities} \label{sec:stats}
Figures \ref{fig:1p} and \ref{fig:hist_masses} show that scattering can account for the moderate ($e \gtrsim 0.1$) to high ($e \sim 0.9$) eccentricities of solitary WJs, but fails to reproduce the peak at low eccentricities.  This may indicate two populations of WJs.  Here we present a mixture model for the solitary WJ eccentricities, augmenting our sample of one-planet systems obtained from \fiducial\ with a population of low-eccentricity planets (that did not undergo scattering).  For simplicity, we neglect the uncertainties of observed eccentricities.  

Consider a mixture model such that the probability density function for eccentricities takes the form
\be
f(e) = \mathcal{F} f_{\rm circ} (e)+ (1 - \mathcal{F}) f_{\rm scat}(e),
\label{eq:pdf}
\ee
where $f_{\rm circ}$ is the injected probability density function of low-eccentricity planets, and $f_{\rm scat}$ is that obtained from \fiducial.  The free parameter $\mathcal{F} = [0,1]$ quantifies the  low-eccentricity planet fraction.

\begin{figure}
\centering 
\includegraphics[scale=0.6]{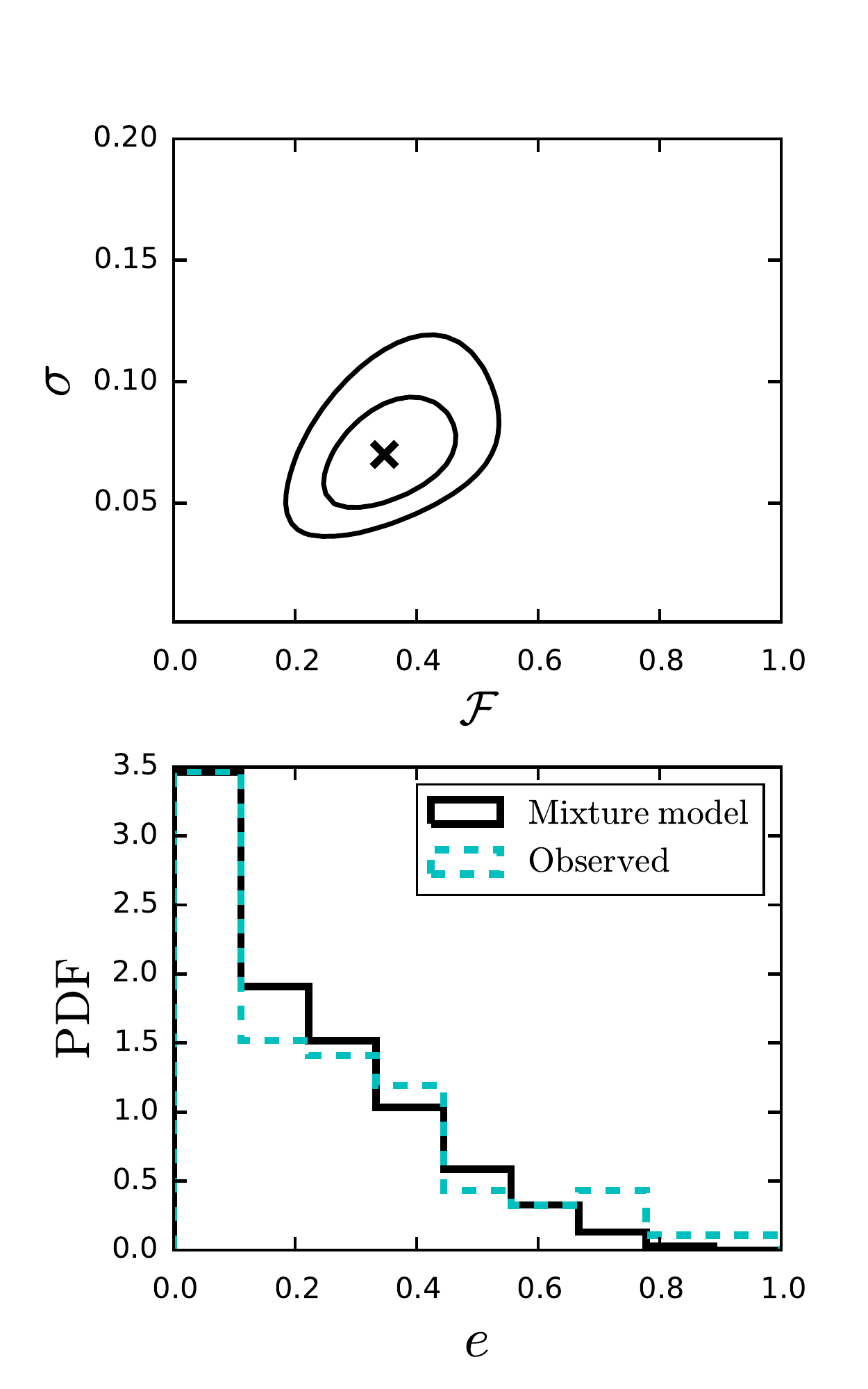}
\caption{{\it Top:} Estimated parameters of the mixture model discussed in Section \ref{sec:stats} (see equation[\ref{eq:pdf}]).  The cross indicates the maximum likelihood estimate (MLE), and the contours indicate the $68 \%$ and  $95 \%$ confidence intervals. {\it Bottom:} Resulting eccentricity distribution from the mixture model using the MLE values $\mathcal{F} = 0.35$ and $\sigma = 0.07$, along with the distribution of observed solitary WJs.}
\label{fig:mle}
\end{figure}

We adopt a half-Gaussian for $f_{\rm circ}(e)$ peaked at $e = 0$ and truncated at $e = 1$, and explore a range of characteristic widths $\sigma$, in the range $[0-0.2]$. This accounts for the fact that modest eccentricities may be generated by factors other than scattering, e.g. by planet-disk interactions.  To construct a smooth function for $f_{\rm scat}(e)$, we apply a Gaussian Kernel-density estimate to our simulated eccentricities.  Armed with $f_{\rm circ}(e)$ and $f_{\rm scat}(e)$, we explore a grid of values for the parameters $(\mathcal{F},\sigma)$, and calculate the likelihood function for obtaining the observed eccentricities $e_i$,
\be
\mathcal{L} =  \displaystyle \prod_{i=1}^{n_{\rm obs}}  \bigg[\mathcal{F} f_{\rm circ} (e_i)+ (1 - \mathcal{F}) f_{\rm scat}(e_i) \bigg],
\ee
with $n_{\rm obs} = 83$. Maximizing $\mathcal{L}$ yields the parameter estimates $\mathcal{F} \approx 0.35$ and $\sigma \approx 0.07$.  This estimate, alongside $68 \%$ and $95 \%$ contours is shown in Fig.~\ref{fig:mle}.  Taking the $95 \%$ contour as the uncertainty, we find that $\mathcal{F}$ lies in the range $0.18 -  0.54$.  

In other words, the observed eccentricity distribution of solitary WJs is consistent with scattering having occurred in roughly half or more of systems.  This finding also holds for the \fourp\ scattering simulations: Performing an identical analysis for the \fourp\ scattering results, we obtain similar parameter estimates, with $\mathcal{F} \approx 0.39$ and $\sigma \approx 0.07$.

\section{Conclusion}\label{sec:conclusion}
In this paper we have undertaken a comprehensive study of giant planet scattering at sub-AU distances ($0.1-1$ AU) from the host star. Our initial system of multiple giant planets in nearly circular orbits is consistent with WJs that formed either in-situ or by disk migration without substantial eccentricity excitation.  Our goal is to catalogue the scattering pathways and outcomes, and to identify the extent to which scattering may have contributed to the observed population of eccentric WJs. We focus primarily on systems of three initially unstable planets, but also briefly consider systems of four planets.

\subsection{Key Results}
\begin{itemize}

\item For our explored parameter space (see Table \ref{table1}), three-planet scatterings result in roughly equal proportions of one-planet and two-planet systems.  These are produced through a combination of collisions and ejections (see Fig.~\ref{fig:bratio}).  The first (or only) planet ``loss'' is a planet-planet collision $60 \% - 80 \%$ of the time.  For systems that later undergo a second planet loss, ejections occur over $90 \%$ of the time.  Two-planet systems are produced almost entirely from planet-planet collisions, whereas one-planet systems arise from a combination of collisions and ejections. Branching ratios separated into two-planet and one-planet systems are depicted in Tables \ref{table_2p} and \ref{table_1p}.

\item The outcomes of three-planet scattering depend on the semi-major axes
of the initial systems (Fig.~\ref{fig:bratio_ecc_binned}). The fraction of single planet systems produced ($f_{\rm 1p}$) increases from $\sim 38 \% $ to $\sim 60 \%$ as the initial $a_1$ increases from $0.1$ AU to $1$ AU, with a corresponding increase in the average planetary eccentricities.

\item Scattering tends to produce closely spaced two-planet systems.  For systems of initially three unstable planets, $90 \%$ of the surviving two-planet systems have a semi-major axis ratio $\aout / \ain \lesssim 3$.  For systems of initially four unstable planets, $90 \%$ of two-planet systems have $\aout / \ain \lesssim 5$.

\item The combined eccentricity distribution of all WJ systems (consisting of single WJs and the inner planet of the two-planet systems) produced by scattering agrees well with the observed distribution (see the left panel of Fig.~\ref{fig:hist_masses}).  However, splitting the results into one and two-planet systems (see the middle and right panels of Fig.~\ref{fig:hist_masses}), we find some discrepancies between the simulations and observations: (i) Examining the two-planet systems, we find that scattering produces too many low-eccentricity planets, due to the fact that all two-planet systems arose from planet-planet collisions.  This is inconsistent with the observed, much flatter eccentricity distribution.  This discrepancy is alleviated, but not fully resolved by addition of a fourth planet to the initial scattering setup (see Fig.~\ref{fig:3p_4p}). (ii) Examining the one-planet systems produced by scattering and comparing with observed solitary WJs (those lacking detection of a giant planet companion), we find that scattering well reproduces the substantial tail of modest and high eccentricities, but does not reproduce the observed peak at low eccentricities. 

\item The observed eccentricity distribution of solitary WJs can be reproduced by two populations of planets, with roughly $60 \%$ having undergone violent scattering, and the remaining having experienced a quiescent dynamical history (see Section \ref{sec:stats} and Fig.~\ref{fig:mle}).

\item Planet-planet collisions have a range of impact parameters, with grazing collisions dominating over head-on (see Fig.~\ref{fig:collinfo}).

\end{itemize}

\subsection{Discussion}
Our study shows that in-situ scattering of multiple giant planets is effective in producing eccentric WJs.  However, while scattering provides a good match to the entire observed WJ eccentricity distribution (consisting of both one and two-planet systems), it yields an excess of circular two-planet systems, and a dearth of circular one-planet systems (see Fig.~\ref{fig:hist_masses}).  A possible resolution to this discrepancy is if many observed solitary WJs have an undetected outer companion, thereby shifting the excess of circular planets from the two-planet systems to the one-planet systems. However, given the large RV amplitudes of the outer planets obtained from our scattering experiments ($10-100$ m/s, see Fig.~\ref{fig:rvamp}), such companions should be readily detectible.  Thus, observed solitary WJs are unlikely to have enough undetected outer companions to resolve the discrepancy in eccentricity distributions (see Fig.~\ref{fig:ecc_rvcut}).

Besides the disagreement in eccentricity distributions, the spacings of two-planet systems generated through in-situ scattering are also inconsistent with many of the observed two-planet systems.  Over $90 \%$ of the two-planet systems produced from the \fiducial\ simulations have semi-major axis ratio $\aout / \ain \lesssim 3$, whereas many observed systems are more hierarchical.  The tendency for scattering to result in compact two-planet systems was pointed out by \cite{raymond2009}.  Many of the observed eccentric two-planet systems remain challenging to explain: They are too compact to have originated from a traditional high-eccentricity migration scenario \citep{antonini2016}, yet are too hierarchical to have formed via in-situ scattering. Highly eccentric, hierarchical systems (with $\aout / \ain \gtrsim 10$) could in principle have originated from a scattering event which initially occurred at several AU distances or beyond, as explored by \cite{mustill2017}.  A thorough study of the parameter space for initial systems consisting of a single WJ and two or more unstable planets much farther out is an important, but computationally expensive problem. 

Our study (Section \ref{sec:stats}) indicates that a two-population model well reproduces the observed eccentricity distribution of solitary WJs, consisting of a $\sim 60 \%$ scattered component and $\sim 40 \%$ quiescent component. The quiescent component is consistent with the finding by \cite{huang2016} that a substantial fraction ($\sim 50 \%$) of WJs have low-mass neighbors (and therefore have experienced no ``dynamical violence").  Two populations of WJs have been previously proposed by \cite{dawson2013}, consisting of low-eccentricity systems around low-metallicity stars, and higher eccentricity systems around high-metallicity stars.  Scattering is consistent with such a metallicity trend, under the expectation that multiple giant planets on unstable orbits form more easily around higher metallicity stars.  Our success in reproducing many of the WJ eccentricities with planet-planet scattering \citep[see also][]{marzari2019,frelikh2019} indicates that formation of multiple giant planets at sub-AU distances, or arrival at sub-AU orbits due to disk migration in a compact configuration, may be common.  However, extrapolating our results to the HJ population would imply an inconsistency with observations. Planet scattering interior to $\sim 0.1$ AU would result in more planet-planet collisions (see Fig.~\ref{fig:bratio_ecc_binned} and also \citealt[][]{petrovich2014}), leading to an even larger fraction of compact two-planet systems. Since HJs lack close, massive companions, scattering of three or more giant planets within $\sim 0.1$ AU is in contradiction with observations (although scattering of only two planets would remain viable).  This may imply that HJs and WJs often have distinct formation and dynamical histories.

Our three-planet scattering simulations show that similar numbers of one and two-planet systems are produced, while our limited four-planet scattering experiments suggest the resulting one-planet systems could outnumber the two-planet systems by a factor of three. Only a subset of the observed two-planet systems are compact enough to be consistent with the scattering predictions from this paper (7 out of 24 systems have $\aout / \ain < 3$, see also Section \ref{sec:obs}).  Given the observed sample of 83 solitary WJs, the ``scattered + quiescent'' model would predict a somewhat larger number of compact two-planet systems than is currently observed.

Better consistency between the observed relative planet multiplicities and those obtained from the ``scattering + quiescent'' model may be obtained if some solitary WJs have a close giant planet companion, or if some of the simulated two-planet systems later become unstable.  As discussed in Section \ref{sec:ecc}, the RV amplitudes of the companions obtained from scattering are large, so the second explanation may be more favorable.  The relative number of one and two-planet systems will increase with time, as some of the two-planet systems undergo instabilities.  Follow-up long-term integrations of a small subset of the two-planet systems (spanning $10^9$ orbits of the inner planet) indicated that at least $\sim 10 \%$ of the two-planet systems will eventually go unstable, but the actual number may be higher over Gyr timescales. Furthermore, physical effects not included in our scattering calculations (e.g. perturbations from a distant companion) may act to destabilize some of the compact two-planet systems. 

\section*{Acknowledgments}
KRA is grateful to Tom Loredo for assistance with the calculation presented in Section \ref{sec:stats}, and thanks Ruth Murray-Clay, Eric Ford, and Renata Frelikh for useful discussions. We also thank the anonymous referee for insightful feedback and comments.  This work has been supported in part by NASA grants NNX14AG94G and NNX14AP31G, and NSF grant AST-1715246.  KRA and BP were supported by NASA Earth \& Space Science Fellowships in Planetary Science. This research has made use of the Exoplanet Orbit Database and the Exoplanet Data Explorer at exoplanets.org.  Simulations in this paper made use of the \rebound\ code which can be downloaded freely at http://github.com/hannorein/rebound.

\appendix

\section{Treatment of Planet-Planet Collisions} \label{sec:collision}
As in previous N-body works, this paper has assumed completely inelastic collisions, so that once the distance between two planets becomes less than the sum of their physical radii, the planets merge conserving mass and momentum.  Such a treatment is clearly a simplification. In reality, giant planets will tidally interact upon close encounters, possibly leading to mass loss.  In addition, head-on versus grazing collisions may result in different outcomes.  Although expanding on this treatment is beyond the scope of this paper, here we discuss some of the collisional dynamics and results from the \fiducial\ simulations. These results will be useful as initial conditions for future hydrodynamical simulations of planet-planet collisions.

\begin{figure}
\centering 
\includegraphics[scale=0.45]{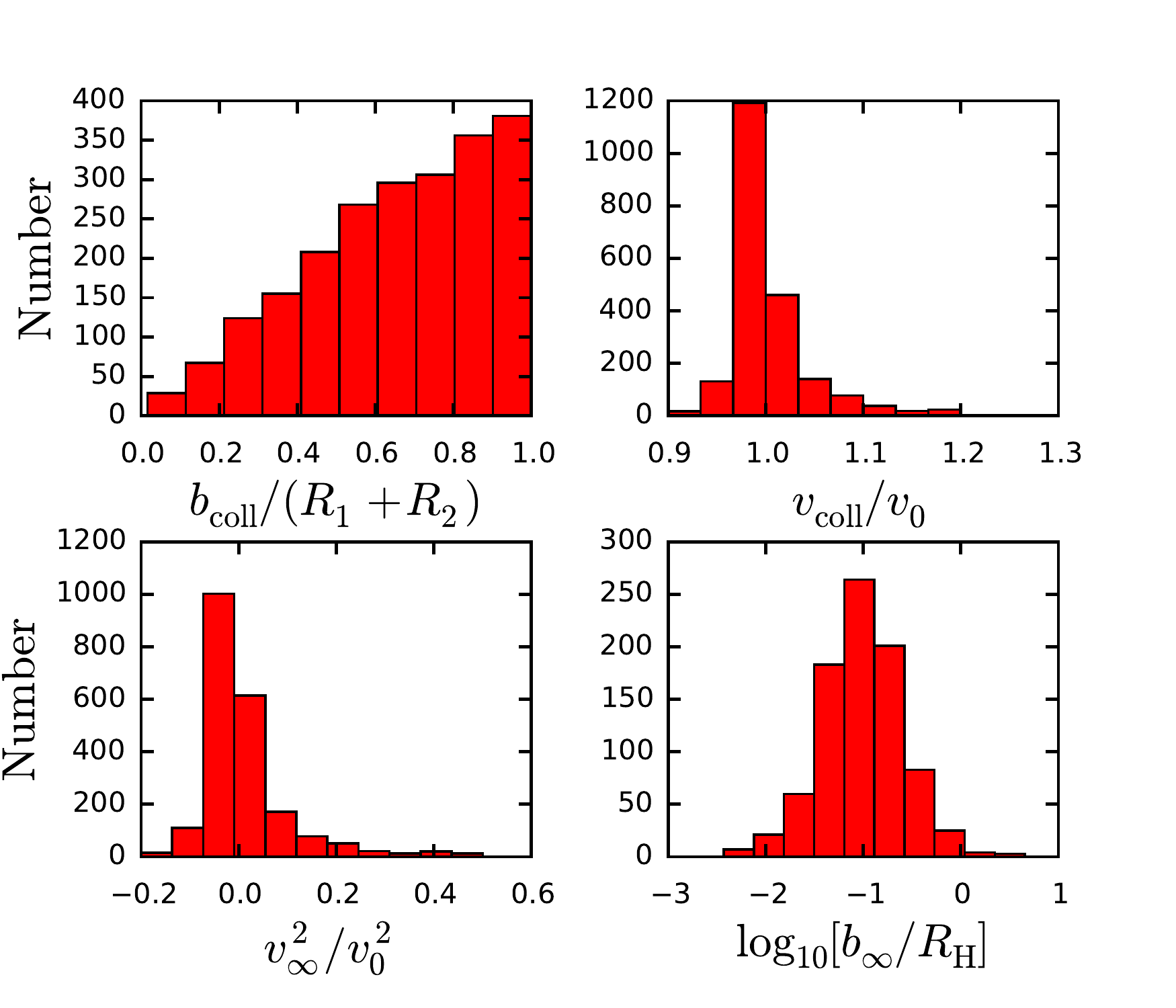}
\caption{Properties of planet-planet collisions from the \fiducial\ simulations (see Appendix \ref{sec:collision}).}
\label{fig:collinfo}
\end{figure}

We define the collision impact parameter, $\bcoll = |{\bf r} \times \hat{{\bf v}}|$ where ${\bf r}$ is the relative distance between the centers of mass of the two planets, and $\hat{{\bf v}}$ is the relative velocity unit vector.  The top left panel of Fig.~\ref{fig:collinfo} shows that many planet-planet collisions are grazing, with $b_{\rm coll} / (R_1 + R_2) \sim 1$, where $R_1$ and $R_2$ are the planetary radii.

Next we discuss some aspects of the collision of two planets $m_1$ and $m_2$.  Once the planet-planet gravitational interaction becomes sufficiently strong so that the tidal gravity from the host star can be ignored, the scattering process up to just before merging can be modeled approximately as an isolated two-body encounter.  Denote $v_{\rm \infty}$ as the relative velocity of two planets ``at infinity'', and $v_{\rm coll}$ as the relative velocity just before collision. Energy conservation yields
\be
v_{\infty}^2 = v_{\rm coll}^2 - v_0^2,
\label{eq:energy_cons}
\ee
where $v_0^2 = 2 G M/(R_1 + R_2)$, with $M$ the sum of the planet masses.  Angular momentum conservations yields
\be
b_{\infty} v_{\infty} = b_{\rm coll} v_{\rm coll},
\ee
where $b_{\infty}$ is the impact parameter at ``infinity''.

The upper right panel of Fig.~\ref{fig:collinfo} shows $v_{\rm coll} / v_0$.  As expected, the majority of collisions have $v_{\rm coll} / v_0 \sim 1$. The lower left panel of Fig.~\ref{fig:collinfo} shows $v_{\infty}^2 / v_0^2$.  Typically $v_{\infty}^2 / v_0^2 \ll 1$. Negative values of $v_{\infty}^2$ indicate that the two planets approach each other on an elliptical orbit.  Finally, the lower right panel of Fig.~\ref{fig:collinfo} shows $b_{\infty}$ in units of the Hill radius, $R_{\rm H} = a (M/3 M_\star)^{1/3}$.  

\bibliography{refs}

\begin{thebibliography}{}
\makeatletter
\relax
\def\mn@urlcharsother{\let\do\@makeother \do\$\do\&\do\#\do\^\do\_\do\%\do\~}
\def\mn@doi{\begingroup\mn@urlcharsother \@ifnextchar [ {\mn@doi@}
  {\mn@doi@[]}}
\def\mn@doi@[#1]#2{\def\@tempa{#1}\ifx\@tempa\@empty \href
  {http://dx.doi.org/#2} {doi:#2}\else \href {http://dx.doi.org/#2} {#1}\fi
  \endgroup}
\def\mn@eprint#1#2{\mn@eprint@#1:#2::\@nil}
\def\mn@eprint@arXiv#1{\href {http://arxiv.org/abs/#1} {{\tt arXiv:#1}}}
\def\mn@eprint@dblp#1{\href {http://dblp.uni-trier.de/rec/bibtex/#1.xml}
  {dblp:#1}}
\def\mn@eprint@#1:#2:#3:#4\@nil{\def\@tempa {#1}\def\@tempb {#2}\def\@tempc
  {#3}\ifx \@tempc \@empty \let \@tempc \@tempb \let \@tempb \@tempa \fi \ifx
  \@tempb \@empty \def\@tempb {arXiv}\fi \@ifundefined
  {mn@eprint@\@tempb}{\@tempb:\@tempc}{\expandafter \expandafter \csname
  mn@eprint@\@tempb\endcsname \expandafter{\@tempc}}}

\bibitem[\protect\citeauthoryear{{Adams} \& {Laughlin}}{{Adams} \&
  {Laughlin}}{2003}]{adams2003}
{Adams} F.~C.,  {Laughlin} G.,  2003, \mn@doi [\icarus]
  {10.1016/S0019-1035(03)00081-2}, \href
  {http://adsabs.harvard.edu/abs/2003Icar..163..290A} {163, 290}

\bibitem[\protect\citeauthoryear{{Anderson} \& {Lai}}{{Anderson} \&
  {Lai}}{2017}]{anderson2017}
{Anderson} K.~R.,  {Lai} D.,  2017, \mn@doi [\mnras] {10.1093/mnras/stx2250},
  \href {http://adsabs.harvard.edu/abs/2017MNRAS.472.3692A} {472, 3692}

\bibitem[\protect\citeauthoryear{{Anderson}, {Storch}  \& {Lai}}{{Anderson}
  et~al.}{2016}]{anderson2016}
{Anderson} K.~R.,  {Storch} N.~I.,   {Lai} D.,  2016, \mn@doi [\mnras]
  {10.1093/mnras/stv2906}, \href
  {http://adsabs.harvard.edu/abs/2016MNRAS.456.3671A} {456, 3671}

\bibitem[\protect\citeauthoryear{{Antonini}, {Hamers}  \&
  {Lithwick}}{{Antonini} et~al.}{2016}]{antonini2016}
{Antonini} F.,  {Hamers} A.~S.,   {Lithwick} Y.,  2016, \mn@doi [\aj]
  {10.3847/0004-6256/152/6/174}, \href
  {http://adsabs.harvard.edu/abs/2016AJ....152..174A} {152, 174}

\bibitem[\protect\citeauthoryear{{Baruteau} et~al.,}{{Baruteau}
  et~al.}{2014}]{baruteau2014}
{Baruteau} C.,  et~al., 2014, \mn@doi [Protostars and Planets VI]
  {10.2458/azu_uapress_9780816531240-ch029}, \href
  {http://adsabs.harvard.edu/abs/2014prpl.conf..667B} {pp 667--689}

\bibitem[\protect\citeauthoryear{{Batygin}, {Bodenheimer}  \&
  {Laughlin}}{{Batygin} et~al.}{2016}]{batygin2016}
{Batygin} K.,  {Bodenheimer} P.~H.,   {Laughlin} G.~P.,  2016, \mn@doi [\apj]
  {10.3847/0004-637X/829/2/114}, \href
  {https://ui.adsabs.harvard.edu/abs/2016ApJ...829..114B} {829, 114}

\bibitem[\protect\citeauthoryear{{Beaug{\'e}} \& {Nesvorn{\'y}}}{{Beaug{\'e}}
  \& {Nesvorn{\'y}}}{2012}]{beauge2012}
{Beaug{\'e}} C.,  {Nesvorn{\'y}} D.,  2012, \mn@doi [\apj]
  {10.1088/0004-637X/751/2/119}, \href
  {http://adsabs.harvard.edu/abs/2012ApJ...751..119B} {751, 119}

\bibitem[\protect\citeauthoryear{{Boley}, {Granados Contreras}  \&
  {Gladman}}{{Boley} et~al.}{2016}]{boley2016}
{Boley} A.~C.,  {Granados Contreras} A.~P.,   {Gladman} B.,  2016, \mn@doi
  [\apjl] {10.3847/2041-8205/817/2/L17}, \href
  {https://ui.adsabs.harvard.edu/abs/2016ApJ...817L..17B} {817, L17}

\bibitem[\protect\citeauthoryear{{Chambers}, {Wetherill}  \& {Boss}}{{Chambers}
  et~al.}{1996}]{chambers1996}
{Chambers} J.~E.,  {Wetherill} G.~W.,   {Boss} A.~P.,  1996, \mn@doi [\icarus]
  {10.1006/icar.1996.0019}, \href
  {http://adsabs.harvard.edu/abs/1996Icar..119..261C} {119, 261}

\bibitem[\protect\citeauthoryear{{Chatterjee}, {Ford}, {Matsumura}  \&
  {Rasio}}{{Chatterjee} et~al.}{2008}]{chatterjee2008}
{Chatterjee} S.,  {Ford} E.~B.,  {Matsumura} S.,   {Rasio} F.~A.,  2008,
  \mn@doi [\apj] {10.1086/590227}, \href
  {http://adsabs.harvard.edu/abs/2008ApJ...686..580C} {686, 580}

\bibitem[\protect\citeauthoryear{{Correia}, {Laskar}, {Farago}  \&
  {Bou{\'e}}}{{Correia} et~al.}{2011}]{correia2011}
{Correia} A. C.~M.,  {Laskar} J.,  {Farago} F.,   {Bou{\'e}} G.,  2011, \mn@doi
  [Celestial Mechanics and Dynamical Astronomy] {10.1007/s10569-011-9368-9},
  \href {https://ui.adsabs.harvard.edu/abs/2011CeMDA.111..105C} {111, 105}

\bibitem[\protect\citeauthoryear{{Cumming}, {Butler}, {Marcy}, {Vogt}, {Wright}
   \& {Fischer}}{{Cumming} et~al.}{2008}]{cumming2008}
{Cumming} A.,  {Butler} R.~P.,  {Marcy} G.~W.,  {Vogt} S.~S.,  {Wright} J.~T.,
   {Fischer} D.~A.,  2008, \mn@doi [\pasp] {10.1086/588487}, \href
  {https://ui.adsabs.harvard.edu/abs/2008PASP..120..531C} {120, 531}

\bibitem[\protect\citeauthoryear{{Dawson} \& {Chiang}}{{Dawson} \&
  {Chiang}}{2014}]{dawson2014}
{Dawson} R.~I.,  {Chiang} E.,  2014, \mn@doi [Science]
  {10.1126/science.1256943}, \href
  {http://adsabs.harvard.edu/abs/2014Sci...346..212D} {346, 212}

\bibitem[\protect\citeauthoryear{{Dawson} \& {Johnson}}{{Dawson} \&
  {Johnson}}{2018}]{dawson2018}
{Dawson} R.~I.,  {Johnson} J.~A.,  2018, \mn@doi [\araa]
  {10.1146/annurev-astro-081817-051853}, \href
  {http://adsabs.harvard.edu/abs/2018ARA%26A..56..175D} {56, 175}

\bibitem[\protect\citeauthoryear{{Dawson} \& {Murray-Clay}}{{Dawson} \&
  {Murray-Clay}}{2013}]{dawson2013}
{Dawson} R.~I.,  {Murray-Clay} R.~A.,  2013, \mn@doi [\apjl]
  {10.1088/2041-8205/767/2/L24}, \href
  {http://adsabs.harvard.edu/abs/2013ApJ...767L..24D} {767, L24}

\bibitem[\protect\citeauthoryear{{Dawson} et~al.,}{{Dawson}
  et~al.}{2014}]{dawson2014kep419}
{Dawson} R.~I.,  et~al., 2014, \mn@doi [\apj] {10.1088/0004-637X/791/2/89},
  \href {http://adsabs.harvard.edu/abs/2014ApJ...791...89D} {791, 89}

\bibitem[\protect\citeauthoryear{{Dong}, {Katz}  \& {Socrates}}{{Dong}
  et~al.}{2014}]{dong2014}
{Dong} S.,  {Katz} B.,   {Socrates} A.,  2014, \mn@doi [\apjl]
  {10.1088/2041-8205/781/1/L5}, \href
  {http://adsabs.harvard.edu/abs/2014ApJ...781L...5D} {781, L5}

\bibitem[\protect\citeauthoryear{{Duffell} \& {Chiang}}{{Duffell} \&
  {Chiang}}{2015}]{duffell2015}
{Duffell} P.~C.,  {Chiang} E.,  2015, \mn@doi [\apj]
  {10.1088/0004-637X/812/2/94}, \href
  {http://adsabs.harvard.edu/abs/2015ApJ...812...94D} {812, 94}

\bibitem[\protect\citeauthoryear{{Fabrycky} \& {Tremaine}}{{Fabrycky} \&
  {Tremaine}}{2007}]{fabrycky2007}
{Fabrycky} D.,  {Tremaine} S.,  2007, \mn@doi [\apj] {10.1086/521702}, \href
  {http://adsabs.harvard.edu/abs/2007ApJ...669.1298F} {669, 1298}

\bibitem[\protect\citeauthoryear{{Ford} \& {Rasio}}{{Ford} \&
  {Rasio}}{2008}]{ford2008}
{Ford} E.~B.,  {Rasio} F.~A.,  2008, \mn@doi [\apj] {10.1086/590926}, \href
  {http://adsabs.harvard.edu/abs/2008ApJ...686..621F} {686, 621}

\bibitem[\protect\citeauthoryear{{Ford}, {Havlickova}  \& {Rasio}}{{Ford}
  et~al.}{2001}]{ford2001}
{Ford} E.~B.,  {Havlickova} M.,   {Rasio} F.~A.,  2001, \mn@doi [\icarus]
  {10.1006/icar.2001.6588}, \href
  {http://adsabs.harvard.edu/abs/2001Icar..150..303F} {150, 303}

\bibitem[\protect\citeauthoryear{{Frelikh}, {Jang}, {Murray-Clay}  \&
  {Petrovich}}{{Frelikh} et~al.}{2019}]{frelikh2019}
{Frelikh} R.,  {Jang} H.,  {Murray-Clay} R.~A.,   {Petrovich} C.,  2019, arXiv
  e-prints, \href {https://ui.adsabs.harvard.edu/abs/2019arXiv190603266F} {p.
  arXiv:1906.03266}

\bibitem[\protect\citeauthoryear{{Goldreich} \& {Sari}}{{Goldreich} \&
  {Sari}}{2003}]{goldreich2003}
{Goldreich} P.,  {Sari} R.,  2003, \mn@doi [\apj] {10.1086/346202}, \href
  {http://adsabs.harvard.edu/abs/2003ApJ...585.1024G} {585, 1024}

\bibitem[\protect\citeauthoryear{{Hamers}, {Antonini}, {Lithwick}, {Perets}  \&
  {Portegies Zwart}}{{Hamers} et~al.}{2017}]{hamers2017}
{Hamers} A.~S.,  {Antonini} F.,  {Lithwick} Y.,  {Perets} H.~B.,   {Portegies
  Zwart} S.~F.,  2017, \mn@doi [\mnras] {10.1093/mnras/stw2370}, \href
  {http://adsabs.harvard.edu/abs/2017MNRAS.464..688H} {464, 688}

\bibitem[\protect\citeauthoryear{{Huang}, {Wu}  \& {Triaud}}{{Huang}
  et~al.}{2016}]{huang2016}
{Huang} C.,  {Wu} Y.,   {Triaud} A.~H.~M.~J.,  2016, \mn@doi [\apj]
  {10.3847/0004-637X/825/2/98}, \href
  {http://adsabs.harvard.edu/abs/2016ApJ...825...98H} {825, 98}

\bibitem[\protect\citeauthoryear{{Juri{\'c}} \& {Tremaine}}{{Juri{\'c}} \&
  {Tremaine}}{2008}]{juric2008}
{Juri{\'c}} M.,  {Tremaine} S.,  2008, \mn@doi [\apj] {10.1086/590047}, \href
  {http://adsabs.harvard.edu/abs/2008ApJ...686..603J} {686, 603}

\bibitem[\protect\citeauthoryear{{Kley} \& {Nelson}}{{Kley} \&
  {Nelson}}{2012}]{kley2012}
{Kley} W.,  {Nelson} R.~P.,  2012, \mn@doi [\araa]
  {10.1146/annurev-astro-081811-125523}, \href
  {http://adsabs.harvard.edu/abs/2012ARA%26A..50..211K} {50, 211}

\bibitem[\protect\citeauthoryear{{Kozai}}{{Kozai}}{1962}]{kozai1962}
{Kozai} Y.,  1962, \mn@doi [\aj] {10.1086/108790}, \href
  {http://adsabs.harvard.edu/abs/1962AJ.....67..591K} {67, 591}

\bibitem[\protect\citeauthoryear{{Lee} \& {Chiang}}{{Lee} \&
  {Chiang}}{2016}]{lee2016}
{Lee} E.~J.,  {Chiang} E.,  2016, \mn@doi [\apj] {10.3847/0004-637X/817/2/90},
  \href {https://ui.adsabs.harvard.edu/abs/2016ApJ...817...90L} {817, 90}

\bibitem[\protect\citeauthoryear{{Lee}, {Chiang}  \& {Ormel}}{{Lee}
  et~al.}{2014}]{lee2014}
{Lee} E.~J.,  {Chiang} E.,   {Ormel} C.~W.,  2014, \mn@doi [\apj]
  {10.1088/0004-637X/797/2/95}, \href
  {https://ui.adsabs.harvard.edu/abs/2014ApJ...797...95L} {797, 95}

\bibitem[\protect\citeauthoryear{{Lidov}}{{Lidov}}{1962}]{lidov1962}
{Lidov} M.~L.,  1962, \mn@doi [\planss] {10.1016/0032-0633(62)90129-0}, \href
  {http://adsabs.harvard.edu/abs/1962P%26SS....9..719L} {9, 719}

\bibitem[\protect\citeauthoryear{{Lin} \& {Ida}}{{Lin} \&
  {Ida}}{1997}]{lin1997}
{Lin} D.~N.~C.,  {Ida} S.,  1997, \mn@doi [\apj] {10.1086/303738}, \href
  {http://adsabs.harvard.edu/abs/1997ApJ...477..781L} {477, 781}

\bibitem[\protect\citeauthoryear{{Lin}, {Bodenheimer}  \& {Richardson}}{{Lin}
  et~al.}{1996}]{lin1996}
{Lin} D.~N.~C.,  {Bodenheimer} P.,   {Richardson} D.~C.,  1996, \mn@doi [\nat]
  {10.1038/380606a0}, \href {http://adsabs.harvard.edu/abs/1996Natur.380..606L}
  {380, 606}

\bibitem[\protect\citeauthoryear{{Lithwick} \& {Wu}}{{Lithwick} \&
  {Wu}}{2011}]{lithwick2011}
{Lithwick} Y.,  {Wu} Y.,  2011, \mn@doi [\apj] {10.1088/0004-637X/739/1/31},
  \href {http://adsabs.harvard.edu/abs/2011ApJ...739...31L} {739, 31}

\bibitem[\protect\citeauthoryear{{Lithwick} \& {Wu}}{{Lithwick} \&
  {Wu}}{2014}]{lithwick2014}
{Lithwick} Y.,  {Wu} Y.,  2014, \mn@doi [Proceedings of the National Academy of
  Science] {10.1073/pnas.1308261110}, \href
  {http://adsabs.harvard.edu/abs/2014PNAS..11112610L} {111, 12610}

\bibitem[\protect\citeauthoryear{{Mardling} \& {Aarseth}}{{Mardling} \&
  {Aarseth}}{2001}]{mardling2001}
{Mardling} R.~A.,  {Aarseth} S.~J.,  2001, \mn@doi [\mnras]
  {10.1046/j.1365-8711.2001.03974.x}, \href
  {http://adsabs.harvard.edu/abs/2001MNRAS.321..398M} {321, 398}

\bibitem[\protect\citeauthoryear{{Marzari} \& {Nagasawa}}{{Marzari} \&
  {Nagasawa}}{2019}]{marzari2019}
{Marzari} F.,  {Nagasawa} M.,  2019, \mn@doi [Astronomy and Astrophysics]
  {10.1051/0004-6361/201935065}, \href
  {https://ui.adsabs.harvard.edu/abs/2019A&A...625A.121M} {625, A121}

\bibitem[\protect\citeauthoryear{{Mu{\~n}oz}, {Lai}  \& {Liu}}{{Mu{\~n}oz}
  et~al.}{2016}]{munoz2016}
{Mu{\~n}oz} D.~J.,  {Lai} D.,   {Liu} B.,  2016, \mn@doi [\mnras]
  {10.1093/mnras/stw983}, \href
  {http://adsabs.harvard.edu/abs/2016MNRAS.460.1086M} {460, 1086}

\bibitem[\protect\citeauthoryear{{Mustill}, {Davies}  \& {Johansen}}{{Mustill}
  et~al.}{2017}]{mustill2017}
{Mustill} A.~J.,  {Davies} M.~B.,   {Johansen} A.,  2017, \mn@doi [\mnras]
  {10.1093/mnras/stx693}, \href
  {http://adsabs.harvard.edu/abs/2017MNRAS.468.3000M} {468, 3000}

\bibitem[\protect\citeauthoryear{{Nagasawa} \& {Ida}}{{Nagasawa} \&
  {Ida}}{2011}]{nagasawa2011}
{Nagasawa} M.,  {Ida} S.,  2011, \mn@doi [\apj] {10.1088/0004-637X/742/2/72},
  \href {http://adsabs.harvard.edu/abs/2011ApJ...742...72N} {742, 72}

\bibitem[\protect\citeauthoryear{{Nagasawa}, {Ida}  \& {Bessho}}{{Nagasawa}
  et~al.}{2008}]{nagasawa2008}
{Nagasawa} M.,  {Ida} S.,   {Bessho} T.,  2008, \mn@doi [\apj]
  {10.1086/529369}, \href {http://adsabs.harvard.edu/abs/2008ApJ...678..498N}
  {678, 498}

\bibitem[\protect\citeauthoryear{{Naoz}, {Farr}  \& {Rasio}}{{Naoz}
  et~al.}{2012}]{naoz2012}
{Naoz} S.,  {Farr} W.~M.,   {Rasio} F.~A.,  2012, \mn@doi [\apjl]
  {10.1088/2041-8205/754/2/L36}, \href
  {http://adsabs.harvard.edu/abs/2012ApJ...754L..36N} {754, L36}

\bibitem[\protect\citeauthoryear{{Petrovich}}{{Petrovich}}{2015a}]{petrovich2015lk}
{Petrovich} C.,  2015a, \mn@doi [\apj] {10.1088/0004-637X/799/1/27}, \href
  {http://adsabs.harvard.edu/abs/2015ApJ...799...27P} {799, 27}

\bibitem[\protect\citeauthoryear{{Petrovich}}{{Petrovich}}{2015b}]{petrovich2015co}
{Petrovich} C.,  2015b, \mn@doi [\apj] {10.1088/0004-637X/805/1/75}, \href
  {http://adsabs.harvard.edu/abs/2015ApJ...805...75P} {805, 75}

\bibitem[\protect\citeauthoryear{{Petrovich}}{{Petrovich}}{2015c}]{petrovich20152p}
{Petrovich} C.,  2015c, \mn@doi [\apj] {10.1088/0004-637X/808/2/120}, \href
  {http://adsabs.harvard.edu/abs/2015ApJ...808..120P} {808, 120}

\bibitem[\protect\citeauthoryear{{Petrovich} \& {Tremaine}}{{Petrovich} \&
  {Tremaine}}{2016}]{petrovich2016}
{Petrovich} C.,  {Tremaine} S.,  2016, \mn@doi [\apj]
  {10.3847/0004-637X/829/2/132}, \href
  {http://adsabs.harvard.edu/abs/2016ApJ...829..132P} {829, 132}

\bibitem[\protect\citeauthoryear{{Petrovich}, {Tremaine}  \&
  {Rafikov}}{{Petrovich} et~al.}{2014}]{petrovich2014}
{Petrovich} C.,  {Tremaine} S.,   {Rafikov} R.,  2014, \mn@doi [\apj]
  {10.1088/0004-637X/786/2/101}, \href
  {http://adsabs.harvard.edu/abs/2014ApJ...786..101P} {786, 101}

\bibitem[\protect\citeauthoryear{{Petrovich}, {Wu}  \& {Ali-Dib}}{{Petrovich}
  et~al.}{2019}]{petrovich2019}
{Petrovich} C.,  {Wu} Y.,   {Ali-Dib} M.,  2019, \mn@doi [\aj]
  {10.3847/1538-3881/aaeed9}, \href
  {http://adsabs.harvard.edu/abs/2019AJ....157....5P} {157, 5}

\bibitem[\protect\citeauthoryear{{Ragusa}, {Rosotti}, {Teyssandier}, {Booth},
  {Clarke}  \& {Lodato}}{{Ragusa} et~al.}{2018}]{ragusa2018}
{Ragusa} E.,  {Rosotti} G.,  {Teyssandier} J.,  {Booth} R.,  {Clarke} C.~J.,
  {Lodato} G.,  2018, \mn@doi [\mnras] {10.1093/mnras/stx3094}, \href
  {https://ui.adsabs.harvard.edu/abs/2018MNRAS.474.4460R} {474, 4460}

\bibitem[\protect\citeauthoryear{{Rasio} \& {Ford}}{{Rasio} \&
  {Ford}}{1996}]{rasio1996}
{Rasio} F.~A.,  {Ford} E.~B.,  1996, \mn@doi [Science]
  {10.1126/science.274.5289.954}, \href
  {http://adsabs.harvard.edu/abs/1996Sci...274..954R} {274, 954}

\bibitem[\protect\citeauthoryear{{Raymond}, {Barnes}, {Veras}, {Armitage},
  {Gorelick}  \& {Greenberg}}{{Raymond} et~al.}{2009}]{raymond2009}
{Raymond} S.~N.,  {Barnes} R.,  {Veras} D.,  {Armitage} P.~J.,  {Gorelick} N.,
   {Greenberg} R.,  2009, \mn@doi [\apjl] {10.1088/0004-637X/696/1/L98}, \href
  {http://adsabs.harvard.edu/abs/2009ApJ...696L..98R} {696, L98}

\bibitem[\protect\citeauthoryear{{Rein} \& {Liu}}{{Rein} \&
  {Liu}}{2012}]{rein2012}
{Rein} H.,  {Liu} S.-F.,  2012, \mn@doi [\aap] {10.1051/0004-6361/201118085},
  \href {http://adsabs.harvard.edu/abs/2012A%26A...537A.128R} {537, A128}

\bibitem[\protect\citeauthoryear{{Rein} \& {Spiegel}}{{Rein} \&
  {Spiegel}}{2015}]{rein2015ias15}
{Rein} H.,  {Spiegel} D.~S.,  2015, \mn@doi [\mnras] {10.1093/mnras/stu2164},
  \href {http://adsabs.harvard.edu/abs/2015MNRAS.446.1424R} {446, 1424}

\bibitem[\protect\citeauthoryear{{Rein} \& {Tamayo}}{{Rein} \&
  {Tamayo}}{2015}]{rein2015whfast}
{Rein} H.,  {Tamayo} D.,  2015, \mn@doi [\mnras] {10.1093/mnras/stv1257}, \href
  {http://adsabs.harvard.edu/abs/2015MNRAS.452..376R} {452, 376}

\bibitem[\protect\citeauthoryear{{Tanaka}, {Takeuchi}  \& {Ward}}{{Tanaka}
  et~al.}{2002}]{tanaka2002}
{Tanaka} H.,  {Takeuchi} T.,   {Ward} W.~R.,  2002, \mn@doi [\apj]
  {10.1086/324713}, \href {http://adsabs.harvard.edu/abs/2002ApJ...565.1257T}
  {565, 1257}

\bibitem[\protect\citeauthoryear{{Teyssandier}, {Lai}  \& {Vick}}{{Teyssandier}
  et~al.}{2019}]{teyssandier2019}
{Teyssandier} J.,  {Lai} D.,   {Vick} M.,  2019, arXiv e-prints, \href
  {http://adsabs.harvard.edu/abs/2019arXiv190105006T} {}

\bibitem[\protect\citeauthoryear{{Tsang}, {Turner}  \& {Cumming}}{{Tsang}
  et~al.}{2014}]{tsang2014}
{Tsang} D.,  {Turner} N.~J.,   {Cumming} A.,  2014, \mn@doi [\apj]
  {10.1088/0004-637X/782/2/113}, \href
  {http://adsabs.harvard.edu/abs/2014ApJ...782..113T} {782, 113}

\bibitem[\protect\citeauthoryear{{Vick}, {Lai}  \& {Anderson}}{{Vick}
  et~al.}{2019}]{vick2019}
{Vick} M.,  {Lai} D.,   {Anderson} K.~R.,  2019, \mn@doi [\mnras]
  {10.1093/mnras/stz354}, \href
  {http://adsabs.harvard.edu/abs/2019MNRAS.484.5645V} {484, 5645}

\bibitem[\protect\citeauthoryear{{Wisdom} \& {Holman}}{{Wisdom} \&
  {Holman}}{1991}]{wisdom1991}
{Wisdom} J.,  {Holman} M.,  1991, \mn@doi [\aj] {10.1086/115978}, \href
  {https://ui.adsabs.harvard.edu/abs/1991AJ....102.1528W} {102, 1528}

\bibitem[\protect\citeauthoryear{{Wu} \& {Murray}}{{Wu} \&
  {Murray}}{2003}]{wu2003}
{Wu} Y.,  {Murray} N.,  2003, \mn@doi [\apj] {10.1086/374598}, \href
  {http://adsabs.harvard.edu/abs/2003ApJ...589..605W} {589, 605}

\makeatother
\end{thebibliography}

\end{document}